\documentclass[10pt,a4paper]{article}
\usepackage[utf8]{inputenc}
\usepackage[toc,page]{appendix}
\usepackage{amsmath}
\usepackage{amsfonts}
\usepackage{amssymb}
\usepackage{natbib}
\usepackage[utf8]{inputenc}    
\usepackage{natbib}
\usepackage[usenames, dvipsnames]{color}
\usepackage[left=2.0 cm,right=2.0 cm,top=2cm,bottom=2cm]{geometry}
\usepackage{color}
\usepackage{epstopdf}
\usepackage{graphicx}
\usepackage{booktabs}
\definecolor{mypink1}{rgb}{0.858, 0.188, 0.478}
\definecolor{mypink2}{RGB}{219, 48, 122}
\definecolor{mypink3}{cmyk}{0, 0.7808, 0.4429, 0.1412}
\definecolor{mygray}{gray}{0.6}
\title{Asymptotics of work distributions in a stochastically driven system}
\author{Sreekanth K Manikandan\footnote{sreekanth.km@fysik.su.se} and Supriya Krishnamurthy\footnote{supriya@fysik.su.se},\\
\small{\em Department of Physics, Stockholm University}, \\
\small{\em SE-106 91 Stockholm, Sweden\vspace{5pt}}}
\begin{document}
\maketitle
\begin{abstract}
We determine the asymptotic forms of work distributions at arbitrary times $T$, in a class of driven stochastic systems using a theory developed by Engel and Nickelsen (EN theory)   \cite{Nickelsen2011}, which is based on the contraction principle of large deviation theory. In this paper, we extend the theory, previously applied in the context of deterministically driven systems, to a model in which the driving is stochastic. The models we study are described by overdamped Langevin equations and the work distributions in path integral form, are characterised by having quadratic augmented actions. We first illustrate EN theory, for a deterministically driven system - the breathing parabola model, and show that within its framework, the Crooks fluctuation theorem manifests itself as a reflection symmetry property of a certain characteristic polynomial, which also determines the exact moment-generating-function at arbitrary times. We then extend our analysis to a stochastically driven system, studied in \cite{expt,Sabhapandit,Varley}, for both equilibrium and non - equilibrium steady state initial distributions. In both cases we obtain new analytic solutions for the asymptotic forms of (dissipated) work distributions at arbitrary $T$.   For dissipated work in the steady state, we compare the large $T$ asymptotic behaviour of our solution to the functional form obtained in   \cite{Varley}. In all  cases, special emphasis is placed on the computation of the pre-exponential factor and the results show excellent agreement with numerical simulations. Our solutions are exact in the low noise ( $ \beta \to \infty $ ) limit.
\end{abstract}
\section{Introduction}
\label{intro}
Stochastic thermodynamics extends the definition of thermodynamic quantities such as entropy, heat and work, to the level of stochastic trajectories, and has become an important area of research in non-equilibrium statistical mechanics of small systems   \cite{Seifert}. Non-equilibrium fluctuations are particularly relevant for systems with a small number of degrees of freedom, where there are large fluctuations around the average and considerable contributions from rare events. Hence a major advancement in the field was due to the discovery of fluctuation theorems (FTs) which extend equilibrium fluctuation-dissipation relations to far-from-equilibrium regimes. Depending on the initial conditions of the system under consideration, FTs put constraints on various thermodynamic distributions and demonstrate that the second law holds statistically in systems with stochastic dynamics; a positive entropy production is exponentially more likely than its negative counterpart. Among the various versions of the FT, the notable ones are the Crooks Fluctuation Theorem (CFT)  \cite{Crooks}, and the Jarzynski Equality (JE)  \cite{Jarzynski1, Jarzynski2} which relate non-equilibrium quantities such as the thermodynamic (Jarzynski) work to equilibrium quantities such as the free energy difference. Advancements in technology have allowed experimental verifications of these fluctuation theorems in a variety of systems (see references in    \cite{Seifert}).\par
Relatedly, there has been significant interest in calculating work distributions for different models. A well studied example is that of a colloidal particle in a harmonic optical trap, where either the mean position of the trap   \cite{Vanzone1, Vanzone2, van3, van4} or the stiffness   \cite{speck,Nickelsen2011,bretpjar,bretpprl,Park} is externally modulated. In the literature these models are referred to as the \textit{sliding parabola} or the \textit{breathing parabola} respectively. For the sliding parabola, if the driving is deterministic, the work distribution is known to be  a Gaussian   \cite{Vanzone1}. For the breathing parabola, the solution given in    \cite{speck} is formally exact for an arbitrary driving protocol. In    \cite{loghar}, an exact calculation of the work distribution has been carried out for the Brownian particle in a logarithmic-harmonic potential. In    \cite{Saha2015}, exact work statistics have been obtained for a Brownian particle in the presence of non-conservative forces such as torques. \par
It is however hard to find an exact expression for the full work distribution $P(W)$ except in the few cases mentioned above. This problem is resolved to a certain extent by using techniques based on large deviation theory    \cite{Touchette}. One of the recent developments in this regard is a theory developed in    \cite{Nickelsen2011} by Engel and Nickelsen (EN Theory), which is used to compute the tail forms of work distributions including the pre-exponential factor at arbitrary times $T$. To derive the asymptotic probability for a certain rare work value, the probability of the most likely trajectory that gives rise to this specific work value is considered. The constraint
to rare work values is effectively equivalent to a low temperature limit. The advantage of this method over other similar large deviation techniques    \cite{sabha1,Sabhapandit} is that the results for the tails are also valid for very small time durations of the process which are typical for experimental situations. EN theory also has the additional benefit that it reduces the problem of computing the asymptotic form of the work distributions to solving a system of ordinary differential equations. One can therefore use available BVP solvers   \cite{bvp} to obtain the full asymptotic form   \cite{Nickelsen2011,Holubec}. In addition to the problems studied in    \cite{Nickelsen2011}, a few more systems have been studies using EN theory   \cite{Nickelsen2,Ryabov}. In    \cite{Holubec}, Holubec {\it et al} put forward a functional-form conjecture to classify the exact asymptotic form of $P(W)$ based on the form of the driving used. In addition, they also obtain the analytic solution to the asymptotic form of $P(W)$ in an absolute value potential (V-potential). Table \ref{tab: 0.} is adapted from    \cite{Holubec}, where the so far known results for the asymptotic forms of work distributions for various driving protocols are listed. In all the above cases the driving protocol $\lambda(t)$ considered, is a deterministic function of time.\par 
\begin{table}
\centering
  \label{tab: 0.}
  \begin{tabular}{ccc}
    \toprule
    Potential U(x(t),\;$\lambda(t)$)& Tail behaviour\\
    \midrule
    \vspace{2mm}
    $\frac{\kappa}{2}\left[x(t)-\lambda(t)\right]^2$ & $A\;e^{-(B\;W-C)^2}$\\\vspace{2mm}
    $\frac{1}{2}\lambda(t)x(t)^2$ & $A \; \vert W \vert^{-1/2} \;e^{-B \vert W \vert} $\\\vspace{2mm}
    $-g\; \text{log }\vert x \vert + \frac{1}{2}\lambda(t)x(t)^2$ & $A \; \vert W \vert^{-(1-\beta \;g)/2} \;e^{-B \vert W \vert} $\\\vspace{2mm}
    $\lambda(t)\; \vert x(t) \vert $ & $A\; e^{-B \;\vert W\vert} $\\
    \bottomrule
       \end{tabular}
       \caption{ \textbf{Asymptotic behaviour of work distributions.} List of asymptotic forms of $P(W)$ known so far adapted from    \cite{Holubec}. $A,B,C$ are constants that depend upon the explicit form of the driving protocol and the duration of the protocol $T$.}
       \end{table}\par       
       In this paper, we will first revisit a model with a deterministic driving protocol, the breathing parabola     \cite{Nickelsen2011}, to familiarize the reader with the methods of EN theory. Then, in this simple model, by considering a particular forward and reverse protocol, we will illustrate the fluctuation theorem for the Jarzynski work. We will show that, within the framework of the EN theory, the fluctuation theorem manifests itself as a symmetry property of a certain characteristic polynomial, which also determines the exact moment generating function at finite times. We then extend this analysis to a stochastically driven system ({\it i.e.} $\lambda(t)$ is stochastic). Stochastic driving protocols have been looked at previously in    \cite{sabha1, Sabhapandit,Varley} and recently in    \cite{unc}. In    \cite{sabha1}, $\lambda(t)$ is taken to be a Gaussian random noise, and the large-time asymptotic form of the distribution of work done by the stochastic force $\lambda(t)$ (the product of the stochastic force $\lambda$ times the displacement $dx$) has been computed, and fluctuation theorems analysed. In    \cite{Sabhapandit}, $\lambda(t)$ is considered instead to be the Ornstein Uhlenbeck process. This is also the stochastic protocol that we study in this paper. The model describes the dynamics of a colloidal particle in a harmonic potential whose mean position is stochastically modulated; we call this model the stochastic sliding parabola (SSP) model. The model studied in    \cite{unc} in the context of the recently discussed finite time thermodynamics uncertainty relation, is equivalent to a discrete version of the SSP. In \citep{Sabhapandit}, it  was claimed that the work fluctuation theorem was valid only in certain regions of the parameter space. Subsequently it was shown in    \cite{Varley}, that when the dissipated work, $W_d$ ( defined as the difference between the Jarzynski work, that differs from the definition in    \cite{sabha1, Sabhapandit} by a boundary term, and the equilibrium free energy difference $\Delta F$) is considered, the Crooks fluctuation theorem is satisfied in all regions of the parameter space. In all these cases, the authors used a moment generating function method, and the $P(W)$ obtained is valid only in the large $T$ limit. In this paper, we study the SSP for both equilibrium and non-equilibrium initial conditions using  EN theory. We show that when the initial points are sampled from an equilibrium distribution, the Jarzynski work ($W$) satisfies a transient fluctuation theorem. To our knowledge, this has not been noted before in the context of this model. We  compute the exact forms of the tails of the PDF for different time durations, and compare them with numerical simulations. For the special case when $\lambda(t=0)$ is unconstrained, we are able to obtain the closed asymptotic form of $P(W)$ as a function of $T$. For non-equilibrium steady state initial conditions, we  obtain the asymptotic form of $P(W_d)$ in a similar manner and compare its large-time behaviour with the asymptotic form obtained using the results in    \cite{Varley}. Our comparison shows that, for very large times, the leading-order asymptotic form of both methods match but they disagree in the sub-leading pre-exponential behaviour. We validate all our results using numerical simulations. In all cases, special emphasis is put on the computation of the pre-exponential factor, which involves calculating a \textit{fluctuation determinant}    \cite{feynman,schulman} that is a \textit{functional determinant} ( of a matrix differential operator    \cite{Kirsten,functional} in the case of the SSP). We do this by using a technique developed in    \cite{Kirsten}, which is based on the spectral -$\zeta$ functions of Sturm-Liouville type operators. \par 
The paper is organized as follows. In Section \ref{methods}, we introduce various techniques and methods used in this article and illustrate EN theory in the context of the breathing parabola problem. Particularly, in Section \ref{flc}, we discuss the FT for a specific choice of forward and reverse protocols and obtain the asymptotic forms of $P(W)$ in each case. In Section \ref{slidp} we extend this analysis to the SSP model. Exact asymptotic forms of PDFs of the Jarzynski work ($W$) and the dissipated work ($W_d$) are obtained for equilibrium and non-equilibrium steady state initial conditions in Sections \ref{slidp:tr} and \ref{slip:ss} respectively. In Section \ref{comp}, we compare the asymptotic forms of $P(W_d)$ with the results from    \cite{Varley}. In Section \ref{con} we present our conclusions. Various technical details of the paper including the computation of the pre-exponential factor, are presented in Appendices \textbf{A - E}.
\section{Basic methods}
\label{methods}
For systems modelled using overdamped Langevin equations, EN theory may be used to compute the asymptotic form of the work probability distribution analytically    \cite{Nickelsen2011,engel}. The theory was initially developed for equilibrium initial conditions, and later generalized to non-equilibrium initial conditions as well    \cite{Ryabov}. The method can be summarized as follows: 
\begin{itemize}
\item[1.] Let $W[x(\cdot)]$ be any functional of the stochastic trajectory $x(\cdot)$. Write down $P(W[x(\cdot)]=W)$ as a path integral, with the corresponding action $S[x(\cdot)]$, by constraining the trajectories to have a work value $W$.
\item[2.] The first order approximation of $P(W)$ for rare $W$ is obtained as 
\begin{equation}
P(W) \sim \exp(-\beta \tilde{S}),
\end{equation}
where $\tilde{S}$ is the action $S$ evaluated along the optimal trajectory that minimizes the action, found by solving the corresponding Euler Lagrange equations together with natural boundary conditions. 
\item[3.] An improved estimate is then obtained by including the pre-exponential factor, which also takes into account the contributions from the trajectories lying close to the optimal trajectory. This is done by expanding $S$ to second order in variations around the optimal trajectory and by performing Gaussian integrations over the variations.
\end{itemize}\par
As is usual for Gaussian integrals, the computation of the pre-exponential factor requires the evaluation of the ratio of determinants of certain differential operators (functional determinants). In this article we use the generalization of a method developed in    \cite{Kirsten}, based on the spectral -$\zeta$ function of Sturm-Liouville operators to determine this ratio. In section \ref{brethp}, we illustrate EN theory using the breathing parabola model. This model has been studied by Engel and Nickelsen   \cite{Nickelsen2011} and the full asymptotic form of the work distribution including the pre-exponential factor has been computed. This model serves as a good starting point to the discussions that follow in the paper since the techniques we will use later are the generalizations of the techniques presented here. We will stick to the notations used in    \cite{Nickelsen2011} unless required otherwise. 
\subsection{Illustrative example: The breathing parabola}
\label{brethp}
The breathing parabola potential is given by
\begin{equation}
\label{bretp}
V(x)=\frac{1}{2}\;\lambda(t)\;x(t)^2,
\end{equation}
where the stiffness of the trap $\lambda(t)$ is the driving protocol which is varied deterministically from a value $\lambda(t=0)=\lambda_0$ to $\lambda(t=T)=\lambda_T$ during each realization of the process. The motion of a colloidal particle in this potential can be described by the overdamped Langevin equation, 
\begin{equation}
\label{bretp:langevin}
\dot{x}(t)=-V^\prime(x)+\sqrt{\frac{2}{\beta}}\eta(t),
\end{equation}
where $V^\prime(x)= dV/dx$ and $\beta$ is the inverse temperature defined as $1/(k_B T)$. $\eta(t)$ is a Gaussian white noise, with $\langle \eta(t) \rangle = 0$, and $\langle \eta(t)\; \eta(s) \rangle = \delta(t-s)$. The Jarzynski work done along a trajectory $x(\cdot)$ of this system for  a time interval $\left[0,T\right]$ is defined    \cite{sekimoto1,sekimoto2} as,
\begin{equation}
\label{bretp:work}
W\left[x(\cdot)\right]\equiv\int_0^{T}\;dt \;\frac{\partial V}{\partial \lambda}\;\dot{\lambda}=\frac{1}{2}\int_0^T dt\;\dot{\lambda}(t)\;x(t)^2.
\end{equation}
$W[x(\cdot)]$ being a functional of a stochastic trajectory, is a random variable by itself. Using the Onsager Machlup formalism    \cite{Onsager1,Onsager2}, the probability of $W[x(\cdot)]$ taking a value $W$ can be written down as a path integral    \cite{Nickelsen2011},
\begin{equation}
\label{bretp:pw2}
P(W)=\frac{\textbf{N}}{Z_0}\int dx_0 \int dx_T\;\int\dfrac{dq}{4\pi/\beta}\int_{x(0)=x_0}^{x(T)=x_T}\; \mathcal{D}[x]\;e^{-\beta \; S[\;x,\;q\;]},\;
\end{equation}
where the {\em augmented} action $S$ is given by,
\begin{equation}
\label{bretp:S}
S[\;x,\;q\;]=V_0(x_0)+\int_{0}^{T}dt\;\left(\frac{1}{4}\;[\dot{x}+V'(x)]^2+\frac{iq}{4}\;\dot{\lambda}(t)x(t)^2\;\right)-\frac{iq}{2}W \equiv S_W-\frac{iq}{2}W.
\end{equation}
$\textbf{N}$ is the normalization constant corresponding to mid point discretization in the functional integral and $Z_0$ is the initial equilibrium partition function. EN theory uses the contraction principle of large deviation theory    \cite{Touchette} to calculate the asymptotic behaviour of $P(W)$ for large $W$. This is implemented by first evaluating the integrals in Eq. \eqref{bretp:S} using the saddle point approximation, and finding the optimal trajectory $(\tilde{x}(\cdot),\tilde{q})$ which minimizes $S$. To find the optimal trajectory,  $S$ is studied near the vicinity of a trajectory $\tilde{x}(t)$ and a value $\tilde{q}$ of $q$ by writing $x(t)=\tilde{x}(t)+y(t)$ and $q=\tilde{q}+r$ and by  expanding $S$  to second order\footnote{
Notice that $S$ will not have expansion terms of order $>2$ in $y(\cdot)$ and $r$. This is because of the quadratic form of the action Eq. \eqref{bretp:S}.} in $y(\cdot)$ and $r$ as, 
\begin{equation}
\label{bretp:Sexp}
S[\;x,\;q\;]=\tilde{S}+S_{lin}+S_{quad},
\end{equation} 
where
\begin{subequations}
\begin{align}
\label{bretp:s0}
\tilde{S} &=S[\;\tilde{x}(\cdot),\;\tilde{q}\;],\\[10pt]
\label{bretp:slin}
\begin{split}
S_{lin} &=\frac{\partial \tilde{V}_0}{\partial x_0}y_0-\frac{1}{2}\; [ \;\dot{\tilde{x}}_0+\lambda_0\tilde{x}_0 ]y_0 + \frac{1}{2}\int dt \;\textbf{A}\;\tilde{x} \; y+\frac{1}{2}[\;\dot{\tilde{x}}_T+\lambda_T\tilde{x}_T]\;y_T\\[5pt]
&+\frac{ir}{2}(\;\int dt \; \frac{\partial \tilde{V}}{\partial \lambda}\dot{\lambda} \; - W\;),
\end{split}\\[10pt]
\label{bretp:squad}
\begin{split}
S_{quad}&= \frac{1}{2}\frac{\partial^2 \tilde{V}_0}{\partial x_0^2}y_0^2-\frac{1}{4}(\;\lambda_0y_0+\dot{y}_0\;)y_0+ \frac{1}{4}\int dt \;y\; \textbf{A}\;y+\frac{1}{4}(\;\lambda_T y_T+\dot{y}_T\;)y_T \\[5pt]&+\frac{ir}{2}\int dt \;\bigg(\; 2\;\dot{\lambda}\;\tilde{x}(t)\;y(t)\;\bigg),
\end{split}
\end{align}
\end{subequations}
where \textbf{A} is a second order Sturm-Liouville type differential operator,
\begin{equation}
\label{bretpA}
\textbf{A}=-\frac{d^2}{dt^2}-\left(\left(1-iq\right)\dot{\lambda}-\lambda^2\right).
\end{equation}
\subsubsection{Leading-order behaviour}
\label{ele}
The leading-order asymptotic behaviour of $P(W)$ is obtained by approximating,
\begin{equation}
\label{lf}
P(W)\sim e^{-\beta \tilde{S}},  
\end{equation}
where $\tilde{S}$ is the action evaluated along the optimal trajectory $\tilde{x}(\cdot)$ and $\tilde{q}$, obtained by demanding that $S_{lin}$ vanishes for an arbitrary variation $\left( y(\cdot), r \right)$. This results in the Euler-Lagrange equations,
\begin{equation}
\label{bretp:ele}
\textbf{A}\tilde{x}=-\ddot{\tilde{x}}(t)-\left(\left(1-i\tilde{q}\right)\dot{\lambda}-\lambda^2\right)\tilde{x}=0,
\end{equation}
with boundary conditions,
\begin{equation}
\label{bretp:elebcs}
\dot{\tilde{x}}_0=\lambda_0\tilde{x}_0,\;\;\dot{\tilde{x}}_T=-\lambda_T\tilde{x}_T.
\end{equation}
The above equations constitute a Sturm-Liouville eigenvalue problem with the parameter $i\tilde{q}$. Therefore there are infinitely many values $ i\tilde{q}^{(n)}$ for which a non trivial solution exists, all of which are saddle points of the action $S$. It can be verified that each $i\tilde{q}^{(n)}$ value corresponds to two solutions, $\pm\tilde{x}^{n}(t)$. It is possible to show that all saddle points except the one corresponding to the smallest $i\tilde{q}(\equiv i\tilde{q}^*)$ value are unstable because they fail the Hessian (second derivative) test\footnote{Since the action is quadratic, the Hessian operator in this case, is the same operator as the operator for the optimal trajectory itself. This means that $\tilde{x}^{(n)}$ is an eigenfunction of the Hessian operator with eigenvalue $0$. According to the Courant nodal theorem    \cite{courant}, $\tilde{x}^{(n)}$ has $n$ nodes. Taken together, these two points imply that the Hessian operator corresponding to $i\tilde{q}^{(n)}$ will have $(n-1)$ eigenfunctions with negative eigenvalues. Therefore the saddle points $x^{(n)}$ for $n>1$, are unstable and do not contribute to the asymptote of $P(W)$.}. Therefore they do not contribute to the asymptote of $P(W)$. The term proportional to $r$ in Eq.\ \eqref{bretp:slin} gives us the constraint equation,
\begin{equation}
\label{bretp:constraint}
W=W[\tilde{x}(\cdot)].
\end{equation}
Using Eq.\ \eqref{bretp:ele},  \eqref{bretp:elebcs} and Eq.\ \eqref{bretp:constraint}, it can be shown that along the optimal trajectory,
\begin{equation}
\label{bretpsq}
\tilde{S}= S[\tilde{x},\tilde{q}]=-\frac{i\tilde{q}^*}{2}\;W.
\end{equation}
Therefore using Eq.\ \eqref{lf}, the leading-order asymptotic behaviour of $P(W)$ is determined to be,
\begin{equation}
\label{bretp:lead}
P(W)\sim e^{\; \beta \;\frac{i\tilde{q}^*}{2}\;W}.
\end{equation}
\subsubsection{The pre-exponential factor}
\label{pre}
The next step is to improve this estimate by also taking into account the fluctuations around the optimal trajectory. This is done by retaining $S_{quad}$ (Eq.\ \eqref{bretp:squad}) in the exponent of Eq. \eqref{bretp:pw2} by computing the Gaussian integrals in 
\begin{equation}
\label{bretp:I1}
\textbf{I}:=\int dy_{0} \int dy_{T} \int \frac{dr}{4\pi / \beta}\int_{y_0}^{y_T} Dy(\cdot)\;e^{-\beta S_{quad}}.
\end{equation}
As already mentioned, since the action is quadratic, the fluctuation governing operator \textbf{A} that appear in $S_{quad}$  is the same as the operator for the optimal trajectory. The operator \textbf{A} therefore has a zero-mode which is nothing but the optimal trajectory itself. Writing integration variable $y$ as a series expansion in terms of the normalized eigenfunctions $\phi_n(t)$ of \textbf{A} as, 
\begin{equation}
\label{bretp:y()}
y(\cdot)=\Sigma_n \; c_n \; \phi_n(t)
\end{equation}
and then performing the Gaussian integrals over the expansion parameters $c_n$ and $r$, it can be shown that the zero-mode of \textbf{A} gets omitted naturally and does not cause any problems to the integral in Eq.\ \eqref{bretp:I1}. This gives us a compact expression for the pre-exponential factor,
\begin{equation}
\label{bretpi4}
\textbf{I}=\frac{\textbf{J}}{\sqrt{\frac{4 \pi}{\beta}}} \frac{1}{\sqrt{d_0^2\; \det \textbf{A}^\prime_{\;i\tilde{q}=i\tilde{q}^*\;}}}.
\end{equation}
The notation $\det\textbf{A}^\prime_{\;i\tilde{q}=i\tilde{q}^*\;}$ stands for a determinant omitting the zero-mode. The factor $d_0$ is defined as,
\begin{equation}
\label{bretp:dn}
d_0\equiv \frac{1}{\vert\vert\tilde{x}\vert\vert}\int_0^T\;dt\;\left( \dot{\lambda}(t)\; \tilde{x}^2(t)\;\right),
\end{equation}
where $\tilde{x}(t)$ is the zero-mode and $\vert\vert\tilde{x}\vert\vert$ stands for its norm. \textbf{J} is a factor stemming from the Jacobian of the transformation of integration variables. The value of \textbf{J} for the breathing parabola problem and also a class of similar potentials has been determined in    \cite{Nickelsen2011}, and can be shown to be equal to,
\begin{equation}
\label{bretp:j}
\textbf{J}=\frac{Z_0}{\textbf{N}}\times \sqrt{\det \textbf{A}_{\;iq=0\;}}.
\end{equation}
A derivation of Eq.\ \eqref{bretpi4} and \eqref{bretp:dn} can be found in    \cite{Nickelsen2011}. Now using Eqns. \eqref{bretp:j}, \eqref{bretpi4}, \eqref{bretp:lead} and \eqref{bretp:pw2} we finally obtain,\begin{equation}
\label{bretp:fullpw}
P(W)= 2\times \sqrt{\frac{\beta}{4 \pi\;d_{0}^2}}\times\sqrt{\frac{\det \textbf{A}_{\;iq = 0\;}}{\det \textbf{A}^\prime_{iq=i\tilde{q}^*}}}\times e^{\;\beta\;\frac{i\tilde{q}^*}{2}\; W}\;(1+O(1/\beta)).
\end{equation}
The factor 2 corresponds to the two equipotent saddle points $(i\tilde{q}^*,\pm \tilde{x}(t))$. Computing the pre-exponential factor now reduces to evaluating $d_0$ and the ratio of functional determinants appearing in the expression above. As we will see, computing $d_0$ is rather straightforward. The evaluation of the determinant ratio is however more involved and is carried out using a technique developed in    \citep{Kirsten}. Notice also that in Eq.\ \eqref{bretp:fullpw}, the protocol $\lambda(t)$ is not specified. In the next Section, we apply the method to a specific forward and the corresponding reverse protocol and exactly compute the asymptotic form of $P(W)$ including the pre-exponential factor, for both cases. We also analyse the FT within the framework of EN theory.
\subsection{ Forward and reverse protocols: Fluctuation relation}
\label{flc}
\begin{figure}[htb!]
 \centering
 \includegraphics[scale=.5]{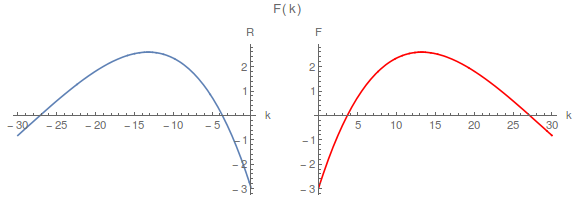}
 \caption{\label{figureone}\textbf{$F(k)$ vs. k for forward (Blue) and reverse (Red) protocols}. The smallest $k$ value for which $F(k)$ vanishes determines the leading-order asymptotic form of $P(W)$ in each case.}
\end{figure}

In this Section, we will compute the exact asymptotic form of the work distribution in the breathing parabola problem, for a specific choice of forward (F) and reverse (R) protocols. The work distributions are then known to satisfy the Crooks fluctuation theorem:
\begin{equation}
\label{bretp:Univ1}
P_F(W)=e^{\beta W-\beta \Delta F} P_R (W).
\end{equation}
We consider the following forward protocol,
\begin{equation}
\label{bretpf}
\lambda(t)=\frac{1}{2-t},\;\, t=0\text{ to }1,
\end{equation}
and the corresponding reverse protocol,
\begin{equation}
\label{bretpr}
\lambda(t)=\frac{1}{1+t},\;\, t=0\text{ to }1.
\end{equation}
This particular reverse protocol was studied as an illustrative example in    \cite{Nickelsen2011}. In order to find the leading-order behaviour of $P(W)$  in each case, one has to obtain the smallest $iq$ values for which the ELE Eqns. \eqref{bretp:ele}, 
\eqref{bretp:elebcs}  have a non trivial solution. Here we use a formalism used in    \cite{Kirsten}. As we will see, this method will turn out to be useful for the computation of the pre-exponential factor as well. First we write the boundary conditions as two matrix equations (for notational simplicity, we will use $x(t)$ instead of $\tilde{x}(t)$.),
\begin{align}
\label{bretpexbc1}
M\;\left[\begin{array}{c}
x_0\\\dot{x}_0
\end{array}\right] &= 0,&N \left[\begin{array}{c}
x_T\\\dot{x}_T
\end{array}\right]&=0.
\end{align}
In this case, M and N are matrices,
\begin{align}
\label{bretpexbc2}
M&=\left[\begin{array}{c c}\lambda_0 & -1 \\0&0 
\end{array}\right],& N &=\left[\begin{array}{cc}0&0\\\lambda_T &1 
\end{array}\right].
\end{align}
It can then be shown that the $iq$ values for which the equations \eqref{bretp:ele}, \eqref{bretp:elebcs} have a non trivial solution, can be obtained as the roots of a function (the characteristic polynomial),
\begin{equation}
\label{bretp:fk}
F(k=1-iq)=\text{Det}\left[M+N H_k(T)\right],
\end{equation}
where $H_k(t)$ is the matrix of fundamental solutions of ELE Eqns. \eqref{bretp:ele},\eqref{bretp:elebcs} defined as     \cite{Kirsten,Kirsten1},
\begin{equation}
H_k(t)=\left[\begin{array}{cc} x_1(t) & x_2 (t)\\
\dot{x}_1(t) & \dot{x}_2(t)
\end{array}
\right].
\end{equation}
We will also make a particular choice of $H_k(t)$, namely that $H_k(0) =\textbf{I}_2$. For convenience, we have defined a variable,
\begin{equation}
\label{bretpk}
k\equiv 1-iq.
\end{equation}
The $i\tilde{q}^*$ value may then be found by looking at the smallest roots of the characteristic polynomial $F(k)$. In Fig. ~\ref{figureone} we plot $F(k)$ vs. $k$ for both the forward and reverse protocols for a time period $T=1$.
\par
Notice the interesting symmetry of the characteristic polynomials namely that, $F(k)$ for the forward protocol is the mirror image of $F(k)$ of the corresponding reverse protocol. As we will show in Appendix \ref{ofk}, this symmetry is a consequence of the FT    \cite{Crooks,Park,linDiff}. As a consequence of this form of $F(k)$, we have
\begin{equation}
\label{bretpksym}
k^*_F =-k^*_R \Rightarrow \left(\frac{i\tilde{q}^*}{2}\right)_F=1-\left(\frac{i\tilde{q}^*}{2}\right)_R.
\end{equation}
For $T=1$, we get $k_F^* = 3.66 = -k_R^*$ and therefore $P_{F/R}(W)$ has the leading-order asymptotic form ($\beta =1$),
\begin{align}
\label{bretpexlead}
P_{F}(W) &\sim e^{-1.33\; \vert W\vert }, &  P_{R}(W)&\sim e^{-2.33\;\vert W \vert}.
\end{align}
Now we move on to the computation of the pre-exponential factor. Obtaining the factor $d_0^2$ is straightforward, using Eq.\ \eqref{bretp:constraint} in Eq. \eqref{bretp:dn} we see that,
\begin{equation}
\label{bretp:d0}
d_0 = \frac{2 W}{\vert \vert x \vert \vert},
\end{equation}
where $\vert \vert x \vert \vert $ is the norm of the zero-mode, and is defined as the usual inner product,
\begin{equation}
\label{bretp:norm}
\vert \vert x \vert \vert^2=\int_0^T dt\;  x^*(t)\; x(t).
\end{equation}
The $*$ in the equation above stands for complex conjugation. The computation of the second factor, which is the ratio of the functional determinants, is rather involved, and may be evaluated using a technique that is developed based on the spectral $\zeta$ function of Sturm-Liouville type operators   \cite{Kirsten}. The power of the method is that it enables one to compute the determinant ratio in terms of the zero-mode itself. In terms of suitably normalized zero-mode solutions, the determinant ratio become
\begin{equation} 
\label{bretp:detratio}
\sqrt{\frac{\det \textbf{A}_{\;iq = 0\;}}{\det \textbf{A}^\prime_{iq=i\tilde{q}^*}}}=\sqrt{
\frac{x_N(T)\times F(1)}{\langle x_N(t) \vert x_N(t) \rangle}}.
\end{equation}
The subscript $N$ denotes that a particular normalization is chosen. Explicit details of the calculation including the choice of normalization will be discussed in Appendix \ref{fdet} and \ref{ec}.\par
Using Eq.\ \eqref{bretp:fullpw}, Eq.\ \eqref{bretp:d0} and Eq.\ \eqref{bretp:detratio}, the exact asymptotic form of $P(W)$ with the pre-exponential factor can now be determined  for both forward and reverse protocols. Doing explicit computations for $T=1$, (see Appendix \ref{ec}) we get,
\begin{align}
\label{bretp:asymform}
P_F(W) &\sim \frac{0.73}{\sqrt{\vert W \vert}} \;e^{-1.33486  \vert W\vert}, & P_R(W) &\sim \sqrt{2}\times \frac{0.73}{\sqrt{\vert W \vert}} \;e^{-2.33486  \vert W\vert}.
\end{align}
The asymptotic forms obtained above are consistent with the Crooks fluctuation relation (Eq.\ \eqref{bretp:Univ1}), with $\beta \Delta F = \frac{1}{2} \text{ ln } \frac{k_f}{k_i} = \frac{1}{2} \text{ ln } \frac{1}{2}$ for the reverse process   \cite{speck,Park}. We have also calculated $P(W)$ numerically, and the results are in good agreement with the theoretical predictions.
\begin{figure}[htb!]
 \centering
 \includegraphics[scale=0.35]{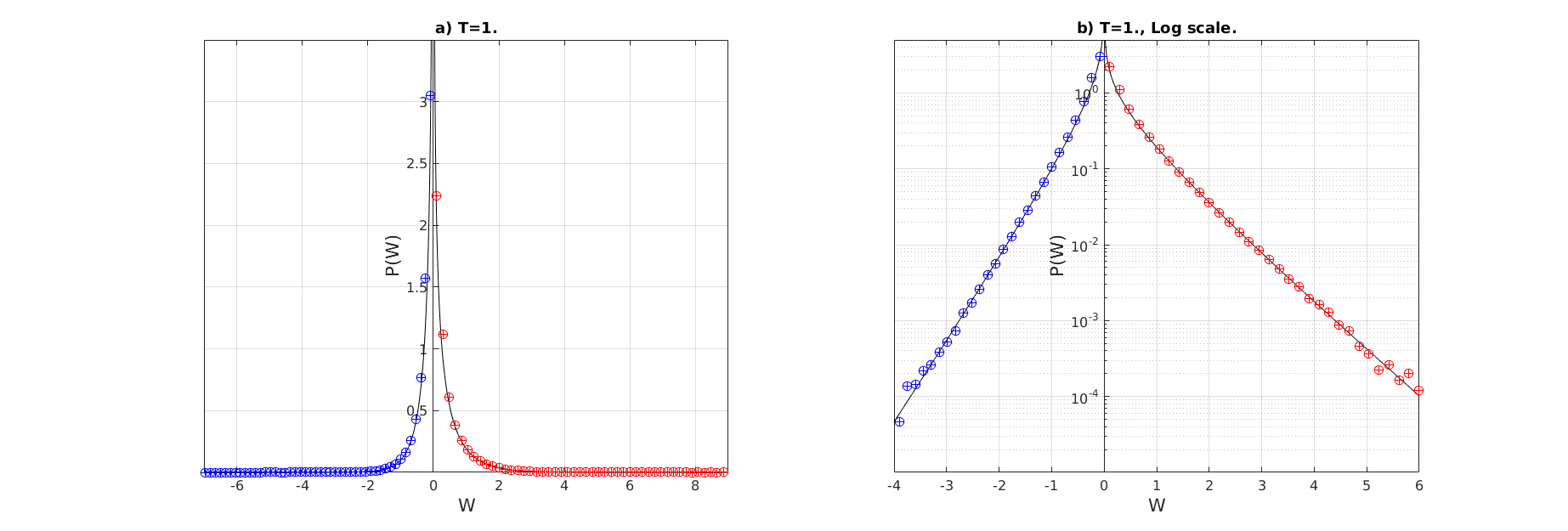}
 \caption{\label{figuretwo}$P(W)$ vs. $W$ for the forward (Red) and reverse (Blue) protocols. The symbols show results from the simulation of the Langevin equation \eqref{bretp:langevin} with a step size of $ \Delta t=0.001$ and an average over $10^6$ realizations. We set $\beta =1$ and $T=1$. The line indicates the asymptotic form including the pre-exponential factor computed using EN theory. Figure 2 b) is a log scale plot of Figure 2 a).}
\end{figure}\par
\subsection{Remarks}
\label{Remarks}
The main points in the calculation of the asymptotic form of $P(W)$ for the breathing parabola hold in general for PDFs with a class\footnote{where all the terms in the action $S_W$ are of degree $2$ in variables $x(t)$ and $\dot{x}(t)$ ( For example, the sliding parabola with deterministic driving \cite{Vanzone1} also has a quadratic action. But the action also contains terms of degree less than 2  in $x(t)$ and $\dot{x}(t)$. As a result of this, the the operator appearing in $S_{quad}$ is different from the one that appear in $S_{lin}$. In this case, EN theory gives the full work distribution which is a Gaussian for all values of $W$.).} of quadratic augmented action $S_W$ (Eq.\ \eqref{bretp:S}), and may be summarised as follows.
\begin{itemize}
\item $P(W)$ has, to leading-order, the functional form,
\begin{equation}
\label{pwleadf}
P(W)\sim \exp(\beta\; \frac{i\tilde{q}^*}{2}\; W),
\end{equation}
where $i\tilde{q}^*$ is the smallest root of some characteristic polynomial $F(k\equiv 1-iq)$. The Crook's fluctuation theorem is seen to manifest itself as a reflection-symmetry property of this characteristic polynomial $F$ around $k=0$. This symmetry is no surprise when we realize that the exact moment generating function of work in this case is given by,
\begin{align}
\label{mgf}
\left\langle e^{-\frac{iq}{2}\; W\left[x(\cdot)\right]}\right\rangle_T &=\sqrt{\frac{F(1)}{F(k)}},& k &\equiv 1-iq.
\end{align}
The $i\tilde{q}^*$ value which determines the asymptotics correspond to the singularity of this moment generating function lying close to zero. We will derive Eq.\ \eqref{mgf} in Appendix \ref{ofk}.\\
\item Including the pre-exponential factor, $P(W)$ takes the form,
\begin{equation}
\label{asymp}
P(W)= 2\times \sqrt{\frac{\beta}{4 \pi\;d_{0}^2}}\times\sqrt{\frac{\det \textbf{A}_{\;iq=0\;}}{\det \textbf{A}^\prime_{iq=i\tilde{q}^*}}}\times e^{\;\beta\;\frac{i\tilde{q}^*}{2}\; W }\;(1+O(1/\beta)),
\end{equation}
where all the factors may be computed in terms of the fundamental solutions of the Euler Lagrange equations. This form for the tail is exact in the low noise limit $\beta\to\infty$, since there are no higher order expansion terms in \eqref{bretp:Sexp} that we are neglecting.\\
\end{itemize}
In the following Sections, we will apply the techniques developed here to a stochastically driven system.     
\section{The stochastic sliding parabola}
\label{slidp}
Consider the dynamics of a colloidal particle in a harmonic trap where the mean position of the trap is externally modulated   \cite{Vanzone1,Vanzone2,engel,Sabhapandit,Varley,sabha1}. Such potentials go by the name \textit{sliding parabola}, and have the general form :
\begin{equation}
\label{slidp:pot }
V(x(t),\lambda(t))=\frac{1}{2}(x(t)-\lambda(t))^2,
\end{equation}
where $x(t)$ is the position variable and $\lambda(t)$ is the externally modulated mean position. We have set the stiffness of the trap to $1$. The dynamics of the colloidal particle in this potential can be described by the Langevin equation,
\begin{equation}
\label{slidp:x}
\dot{x}(t)=-\frac{1}{\tau_r}\frac{\partial V(x,\lambda)}{\partial x(t)}+ \sqrt{2D}\; \eta (t),
\end{equation}
where $\eta(t)$ is a thermal noise and $D$ is the diffusion coefficient. $\eta$ is assumed to be Gaussian with $\langle \eta(t) \rangle = 0$, and $\langle \eta(t)\; \eta(s) \rangle = \delta(t-s)$. One of the natural time scales in the system is given by the relaxation time in the harmonic trap ${{\tau }_{r}}=\gamma /\kappa$ where $\gamma$ is the friction coefficient and $\kappa$ is the stiffness of the trap. In the case of a deterministic driving protocol, the exact statistics of the Jarzynski work done on the colloidal particle is known    \cite{Vanzone1}. The work distribution is a Gaussian and satisfies the transient fluctuation theorem. The sliding parabola with a deterministic driving was also looked at in    \cite{Nickelsen2011,engel}, as a test example for the EN Theory, and in this case the method gives the full probability distribution (not just the large-$W$ form).   Eq.\ \eqref{slidp:x} has also been studied both experimentally    \cite{expt} and analytically    \cite{Sabhapandit,sabha1,Varley} when $\lambda(t)$ is a stochastic driving protocol. One of the cases studied is when $\lambda(t)$ is the Ornstein-Uhlenbeck process given by,
\begin{equation}
\label{slidp:OU}
\dot{\lambda}(t)=-\frac{\lambda(t)}{\tau_0}+\sqrt{2A}\; \xi (t).
\end{equation}
$\xi(t)$ is again assumed to be a Gaussian noise with $\langle \xi \rangle =0$ and $\langle \xi(t)\xi(s) \rangle=\delta(t-s)$. The noise $\xi$ is usually athermal in origin with a diffusion coefficient $A$ as given in Eq.\ \eqref{slidp:OU}. $\tau_0$ gives the second natural time scale in the system in terms of the relaxation time of $\lambda$ correlations. The two noises are assumed to not have cross correlations, {\it i.e.}  $\langle \eta(t)\; \xi(s) \rangle =0$.  Hereafter, we will refer to the coupled equations \eqref{slidp:x} and \eqref{slidp:OU} as the Stochastic Sliding Parabola (SSP).\par
In the remaining Sections of this paper, we study the SSP model for both equilibrium and non-equilibrium steady state initial conditions using EN theory. For equilibrium initial conditions, the dissipation function that satisfies a fluctuation theorem  of the SSP can be identified with the Jarzynski work\footnote{This result can be derived from the ratio of the net probabilities of the forward trajectory and the corresponding time-reversed trajectory as discussed in    \cite{Chernyk}.}. The form of the dissipated work in the steady state has been obtained in    \cite{Varley}. For both situations, the exact form of work distributions at arbitrary times is not known. Here we show that the discussions in Section \ref{Remarks} can be applied to this system, and we can hence compute the exact asymptotic form of both the transient and  steady state work distributions including the pre-exponential factor, at arbitrary times $T$. For steady state initial conditions, we compare our results with    \cite{Varley}, in the appropriate limits. Without loss of generality, for the calculations that follow, we set $A =D=k_B \mathcal{T}$ and $\tau_0=\tau_r=1$.
\subsection{Equilibrium initial condition. : Transient fluctuations}
\label{slidp:tr}
In the first case that we look at, we compute the asymptotic form of the distribution of the Jarzynski work done on the colloidal particle ($W$) by the stochastic force Eq.\ \eqref{slidp:OU} starting from an initial equilibrium distribution given by,
\begin{align}
\label{slidp:pini}
p_{\lambda_0}(x_0)&=\frac{e^{-\beta \; V_0 (x_0,\;\lambda_0)}}{Z_0},&p(\lambda_0)&=\sqrt{\frac{\beta}{2\pi}}\;e^{-\beta\;\frac{\lambda_0^2}{2}},
\end{align}
The particle is assumed be in thermal equilibrium initially for a fixed value of $\lambda_0$ and the partition function $Z_0$ is computed accordingly. The Jarzynski work done on the colloidal particle along each trajectory is defined in the same way as before :
\begin{equation}
\label{slidp:w}
W[x(\cdot),\lambda(\cdot)]=\int_0^{T}\;dt \;\frac{\partial V}{\partial \lambda}\dot{\lambda}.
\end{equation}
In terms of the joint probability density functional of trajectories $\left\lbrace x\left(\cdot\right),\lambda\left(\cdot\right)\right\rbrace_0^T$, the probability density function of work can be written down as,
\begin{equation}
\label{slidp:pw}
P(W)=\frac{\textbf{N}}{Z_0}\sqrt{\frac{\beta}{2\pi}}\int dx_0 \int dx_T \int d\lambda_0 \int d\lambda_T\;\int \frac{dq}{4\pi/\beta}\int_{x_0,\lambda_0}^{x_T,\lambda_T} \;D[x,\lambda]\;e^{-\beta \; S[\;x,\; \lambda,\;q\;]},
\end{equation}
with the action
\begin{equation}
\label{slidp:S}
\begin{split}
S[\;x,\; \lambda,\;q\;]&=\frac{\left(x_0-\lambda_0 \right)^2}{2}+\frac{\lambda_0^2}{2}+\int_{0}^{T}dt\;\bigg(\frac{1}{4}\;[\dot{x}+x-\lambda]^2+\frac{1}{4}\;[\dot{\lambda}+\lambda]^2\\&+ \frac{iq}{2}\;\left(\lambda-x \right)\dot{\lambda}\;\bigg)-\frac{iq}{2}W.
\end{split}
\end{equation}
The normalization constant for this case is    \cite{pathintegral},
\begin{equation}
\label{slidp:norm}
\textbf{N}=\exp\left(\frac{1}{2}\int_0^T\;dt\;\left[\; V^{\prime\prime}(x(t),\lambda(t))+1\;\right]\;\right).
\end{equation}
In order to find the large $W$ asymptotic behaviour of $P(W)$, we will adopt the methods discussed in Section \ref{brethp}. Here that means, we need to identify the optimal choice of both $\tilde{x}(t)$ and $\tilde{\lambda}(t)$ that minimizes the action $S$ for a given value of $W$. We follow the same procedure as before and put $x(t)=\tilde{x}(t)+y(t)$, $\lambda(t)=\tilde{\lambda}(t)+z(t)$ and $q=\tilde{q}+r$ and expand $S$ to second order in $y(\cdot),z(\cdot)$ and $r$.
\begin{equation}
\label{slidp:SExpansn}
S[\;x,\; \lambda,\;q\;]=\tilde{S}+S_{lin}+S_{quad}.
\end{equation}
Notice again that since $S$ is quadratic, the expansion  will not contain terms of order greater than quadratic in $y(\cdot)$, $z(\cdot)$ and $r$.
\subsubsection{Leading-order form of $P(W)$}
\label{slip:lead}
As in the case of the Breathing Parabola problem, The leading-order form of $P(W)$ can be computed in terms of the optimal trajectory $(\tilde{x}(t),\tilde{\lambda}(t))$ as,
\begin{equation}
\label{slidp:leado}
P(W)\sim e^{-\beta\tilde{S}},\; \text{ where }\tilde{S}=S[\;\tilde{x},\; \tilde{\lambda}],
\end{equation}
where $(\tilde{x}(t),\tilde{\lambda}(t))$ solve the following Euler Lagrange equations,
\begin{align}
\label{slidp:ele}
\textbf{A}\left[\begin{array}{c}\tilde{x}(t)\\\tilde{\lambda}(t)
\end{array}\right]&=0,&\text{ where } \textbf{A} &= \left[\begin{array}{cc}-\frac{d^2}{d 
t^2}+1 & k\;\frac{d}{d t}-1\\-k\;\frac{d}{d t}-1 & -\frac{d^2}{d 
t^2}+2 \end{array} \right];\; k\equiv 1-iq,
\end{align}
together with Robin-type boundary conditions,
\begin{subequations}
\label{slidp:elebc}
\begin{alignat}{4}
\tilde{x}(0) -\tilde{\lambda}(0)-\dot{\tilde{x}}(0) &=0,\\
-(k+1)\;\tilde{x}(0) +(k+2)\;\tilde{\lambda}(0)-\dot{\tilde{\lambda}} &=0,\\
\tilde{x}(T)-\tilde{\lambda}(T)+\dot{\tilde{x}}(T) &=0,\\
(k-1)\;\tilde{x}(T)+(2-k)\;\tilde{\lambda}(T)+\dot{\tilde{\lambda}}(T) &=0.
\end{alignat}
\end{subequations}
and the constraint equation,
\begin{equation}
\label{slidp:constraint}
W= W\left[\tilde{x},\tilde{\lambda} \right].
\end{equation}
Notice that (as we have seen in section \ref{ele},) a non trivial solution to the ELEs exists only for some specific values of $iq \equiv i\tilde{q}$, and only the smallest $i\tilde{q}$ value $(\equiv i\tilde{q}^*)$ is relevant. Based on the discussion in Section \ref{Remarks}, we can conclude that, to leading order,
\begin{equation}
\label{slidp:lead1}
P(W) \sim \text{exp}\;(\;\beta\;\frac{i\tilde{q}^*}{2} \;W \;).
\end{equation}
The $ i\tilde{q}^*$ value may be found by looking at the roots of the function,
\begin{align}
\label{slidp:fk}
F(k)&\equiv\text{det}\left[M+NH_k(T)\right], & (k\equiv 1-iq),
\end{align}
corresponding to this problem (see Appendix \ref{ec}). In  Fig. \ref{figurethree} we plot $F(k)$ vs $k$ for $T =1 $.
\begin{figure}[htb!]
 \centering
 \includegraphics[scale=0.5]{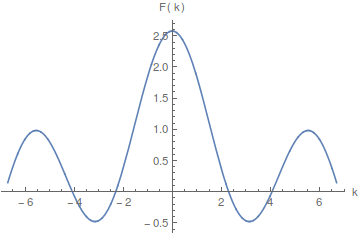}
 \caption{\label{figurethree}\textbf{$F(k)$ vs. k for $T=1$}. The exact asymptotic form of $P(W)$ for positve and negative work values $W$ can be obtained from the smallest positive and negative roots of $F(k)$.}
\end{figure} It may be verified that $F(k)$ is a symmetric function under the transformations $k\rightarrow\; -k$ for any value of $T$. This indicates that the work fluctuations satisfy the fluctuation theorem
\begin{equation}
\label{slidp:Univ1}
P(W)=e^{\beta W} P (-W).
\end{equation}
This form of the Crooks fluctuation theorem is a consequence of the fact that $\Delta F = 0$ for this particular choice of initial conditions.  The smallest roots of $F(k)$ are found to be $k^*=\pm\; 2.300$. Solving Eq.\ \eqref{slidp:ele} with Eq.\ \eqref{slidp:elebc} for $k = \pm\;2.3$, and then using the constraint equation \eqref{slidp:constraint}, it can be verified that $k=+\;2.3$ corresponds to the positive tail of $P(W)$ and $k=-\;2.3$ corresponds to the negative tail of $P(W)$. Hence using Eq.\eqref{slidp:lead1}, the leading-order asymptotic forms of $P(W)$ may be written down as (For $\beta=1$),
\begin{align}
P(W^+) &\sim e^{-0.65\;W}, & P(W^-)&\sim e^{-1.65\vert W \vert}.
\end{align}
Notice that the asymptotic forms are again consistent with the fluctuation theorem (Eq.\ \eqref{slidp:Univ1}).
\subsubsection{$P(W)$ including the pre-exponential factor}
\label{pre2}
As in Section \ref{pre}, the asymptotic estimate for $P(W)$ obtained above can be improved by also taking into account the contributions coming from $S_{quad}$ in Eq.\ \eqref{slidp:SExpansn}. Since the action in Eq.\ \eqref{slidp:S} is quadratic, it can be shown that the fluctuations around the optimal trajectory are again governed by the same operator \textbf{A} (Eq.\ \eqref{slidp:ele}) which determines the optimal trajectory. $P(W)$ including the pre-exponential factor therefore takes the form:
\begin{equation}
\label{slidp:preexp}
P(W)= 2\times \sqrt{\frac{\beta}{4 \pi\;d_{0}^2}}\times\sqrt{\frac{\det \textbf{A}_{\;iq=0\;}}{\det \textbf{A}^\prime_{iq=i\tilde{q}^*}}}\times e^{\;\beta\;\frac{i\tilde{q}^*}{2}\; W }\;(1+O(1/\beta)).
\end{equation}
The form of the Jacobian that is required to derive Eq.\ \eqref{slidp:preexp} is obtained in Appendix \ref{jac}. As in the previous case,  $\det\textbf{A}^\prime$ in Eq.\ \eqref{slidp:preexp} is the determinant of the operator \textbf{A} omitting the zero-mode. $d_0$ is given by,
\begin{equation}
\label{slidp:d0}
d_0=\frac{2 W}{\vert \vert\left[ \begin{array}{c}x\\ \lambda 
\end{array} \right] \vert \vert},
\end{equation}
where $\vert \vert \left[\begin{array}{c}x\\ \lambda 
\end{array} \right] \vert \vert$ is the norm of the zero-mode. The factor 2 in Eq.\ \eqref{slidp:preexp} again accounts for the two equipotent saddle points, $(\pm x(t),\pm\lambda(t),i\tilde{q}^*_\pm)$. Notice that one significant difference from the case of the breathing parabola is that the functional operators appearing in the determinant ratio in Eq.\ \eqref{slidp:preexp} are $2D$ functional operators. We hence generalise the method of functional determinants    \cite{Kirsten} that we used for the $1D$ Sturm-Liouville operator in the previous case, for this $2D$ case, to compute the ratio of the determinants appearing in Eq.\ \eqref{slidp:preexp}. The explicit details are given in Appendices \ref{fdet} and \ref{ec}. For the case $T=1$ one can hence compute,
\begin{equation}
2\times \sqrt{\frac{\beta}{4 \pi\;d_{0,\pm}^2}}\times\sqrt{\frac{\det \textbf{A}^\prime_{\;iq=0\;}}{\det \textbf{A}^\prime_{\;iq=i\tilde{q}^\pm\;}}} =\frac{0.52}{\sqrt{\vert W\vert}},
\end{equation}
and therefore using Eq.\ \eqref{slidp:preexp}, the improved estimate to the positive and negative tails are,
\begin{align}
\label{eq:FF_SSP}
P(W^+) &\sim \frac{0.52}{\sqrt{\vert W\vert}}\; e^{-0.65\;W}, & P(W^-)&\sim\frac{0.52}{\sqrt{\vert W\vert}}\; e^{-1.65\vert W \vert}.
\end{align}
In Table \ref{table:1} we give the exact asymptotic forms of $P(W)$ for different values of $T$.\par
\begin{table}[h!]
  \centering
    \begin{tabular}{ccc}
    \toprule
    T & $P(W^+)$ & $P(W^-)$\\
    \midrule
    \vspace{2mm}
    0.3 & $\frac{0.49}{\sqrt{\vert W \vert}} \;e^{-1.04  \vert W\vert}$ &$\frac{0.49}{\sqrt{\vert W \vert}} \;e^{-2.04  \vert W\vert}$\\\vspace{2mm}
    0.5 & $\frac{0.48}{\sqrt{\vert W \vert}} \;e^{-0.819  \vert W\vert}$ &$\frac{0.48}{\sqrt{\vert W \vert}} \;e^{-1.819  \vert W\vert}$\\\vspace{2mm}
    0.7 & $\frac{0.48}{\sqrt{\vert W \vert}} \;e^{-0.72  \vert W\vert}$ &$\frac{
    0.48}{\sqrt{\vert W \vert}} \;e^{-1.72  \vert W\vert}$\\\vspace{2mm}
    1 & $\frac{0.52}{\sqrt{\vert W\vert}}\; e^{-0.65\;\vert W\vert} $ & $\frac{0.52}{\sqrt{\vert W\vert}}\; e^{-1.65\;\vert W\vert} $\\
    \bottomrule
  \end{tabular}
  \caption{ \textbf{Asymptotic Forms of $P(W,\;T)$}. The exact asymptotic form of $P(W)$ can be determined for any value of $T$. Table \ref{table:1} gives the exactly computed form for a few $T$ values, which may be compared with experiments / numerical simulations.}
  \label{table:1}
\end{table}\par
\begin{figure}[htb!]
 \centering
 \includegraphics[scale=0.35]{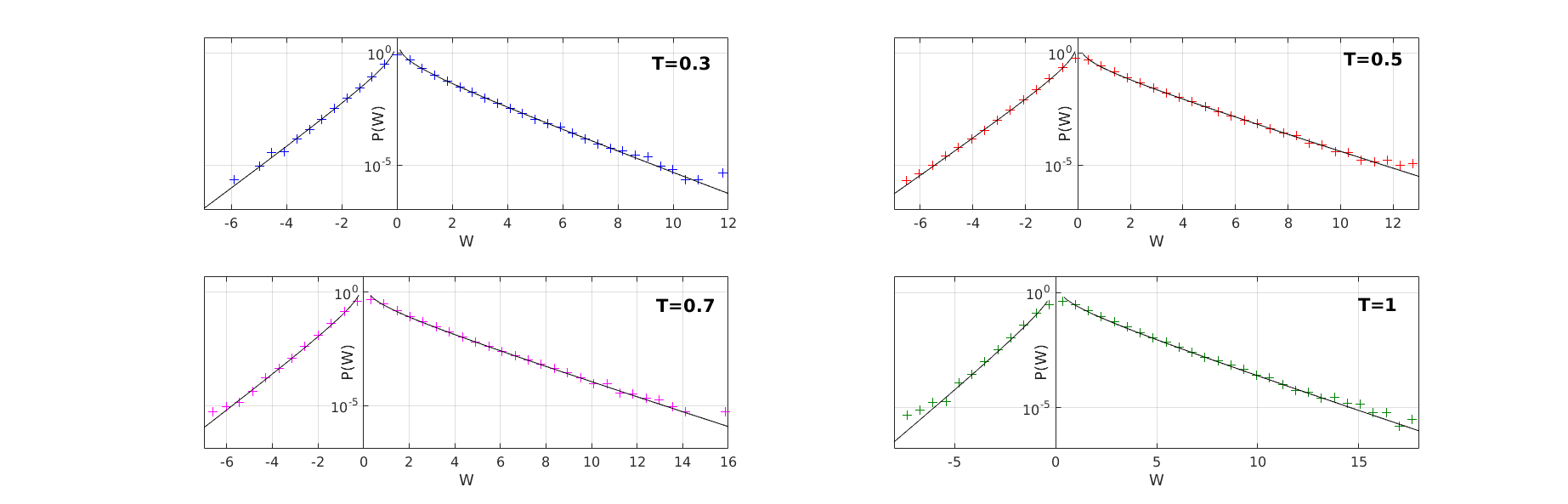}
 \caption{\label{figurefour}\textbf{$P(W)$ vs. $W$; comparison with numerical results}. Symbols represent the results from the numerical simulation of the SSP with a step size of $\Delta t = 0.001$, $\beta =1$ and an average over $10^6$ realizations. Solid lines correspond to the exact forms computed in Table \ref{table:1} using EN theory.}
\end{figure}

We have also numerically integrated the Langevin equations to get an estimate for $P(W)$. For a step size of $\Delta t = 10^{-3}$, averaging over $10^6$ realizations gives a reasonably good agreement with our theoretical predictions. The results are plotted in Figure \ref{figurefour}.
\par
\begin{figure}[htb!]
 \centering
 \includegraphics[scale=0.35]{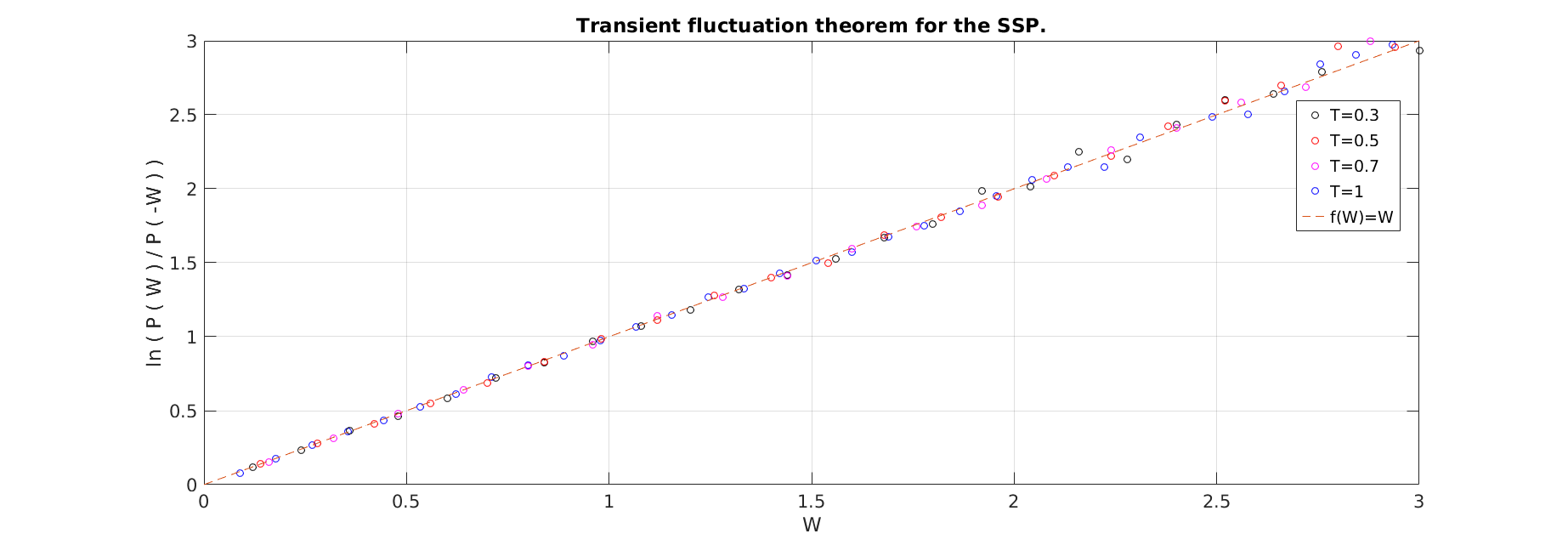}
 \caption{\label{figurefive} \textbf{Verification of the transient fluctuation theorem for different values of $T$.} The symbols corresponds to the simulation data obtained previously. The dashed line correspond to the identity function, $f(W)=W$.}
\end{figure}

The results agree for large values of $W$, except for the very far tails where there are not sufficiently many sample points. In Figure \ref{figurefive} we present the numerical result, verifying the fluctuation theorem.\par
\par
\subsubsection{$\lambda_0$ arbitrary.}
\label{free}
The case when we leave $\lambda_0$ unconstrained,  (equivalent to sampling $\lambda_0$ from a Gaussian distribution with very large variance), is a special case in which one can obtain a closed asymptotic form of $P(W)$ as a function of $T$ - the time period of duration of the protocol. Following similar calculations as for the previous case (Appendix \ref{ec}), but with modified boundary conditions \eqref{slidp:elebc}, we find for the positive tail,
\begin{equation}
P(W^+)\sim 2\times \sqrt{\frac{2 T^2+6 T-7 e^{2 T}+7}{4 T-2 e^{2 T}+2}}\times \sqrt{\frac{\beta}{8 \pi \; \vert W \vert\;}} ,
\end{equation}
and for the negative tail,
\begin{equation}
P(W^-)\sim 2 \times \sqrt{\frac{2 T^2+6 T-7 e^{2 T}+7}{4 T-2 e^{2 T}+2}}\times \sqrt{\frac{\beta}{8 \pi \; \vert W \vert\;}}  \times e^{-\beta \vert W \vert}.
\end{equation}
For $T\rightarrow \infty $ we find that the time-dependent part of the the pre-exponential factor converges to a value $\sqrt{\frac{7}{2}}$. Notice that the positive tail of the probability distribution decays as a power-law in $W$, and therefore the mean and higher moments do not exist. This behaviour of the tails can be attributed to the large fluctuations in the system which contribute to positive work values. However, for any finite variance of the initial Gaussian distribution of $\lambda_0$, the work distributions can be shown to have tails of the form \eqref{eq:FF_SSP}, and well defined moments. 
\subsection{The steady state fluctuations.}
\label{slip:ss}
In this section we study the steady state work fluctuations in the SSP. The functional form of the dissipated work ($W_d$) in the steady state of the SSP was identified in    \cite{Varley}. It was then shown to satisfy the FT by computing the corresponding moment generating function in the large $T$ limit. Here we look at the exact asymptotic form of $P(W_d)$ using EN theory. We will then make a comparison with results from    \cite{Varley} in the appropriate limits.
\subsubsection{Asymptotic form of $P(W_d)$ for large $W_d$.}
The dissipated work in the steady state in the time interval $[0,T]$ for the SSP is given by    \cite{Varley},
\begin{equation}
\label{slidp:ep}
W_d\left[x,\lambda\right]=\int_0^T\;dt\; \lambda(t)\;\dot{x}(t)+\frac{1}{10} \left(\;x(0)^2+4\; x(0)\; \lambda(0)-x(T)^2-4\; x(T)\; \lambda(T)-\lambda(0)^2+\lambda(T)^2\;\right).
\end{equation}
The form of the dissipation function may also be identified from the formalism presented in    \cite{Chernyk}. Sampling the initial points from the stationary probability distribution (from    \cite{Sabhapandit}),
\begin{equation}
\label{slidp:pst}
p_{st}(x(t),\;\lambda(t))=\frac{1}{\sqrt{5} \pi }\;\text{ exp } \left[-\beta \frac{2\; x(t)^2-2\; x(t)\; \lambda(t)+3\; \lambda(t)^2}{5}\;  \right],
\end{equation}
one can write down the probability distribution for dissipated work in the steady state as,
\begin{equation}
\label{slidp:pwd}
P(W_d)= \int dx_0 \;\int d\lambda_0\;\int dx_T\;\int d\lambda_T\;\int\dfrac{dq}{4\pi/\beta}\int_{x(0),\lambda(0)=x_0,\lambda_0}^{x(T),\lambda(T)=x_T,\lambda_T}\; \mathcal{D}[x,\; \lambda]\;e^{-\beta \; S[\;x,\; \lambda,\;q\;]},\;
\end{equation}
with the augmented action
\begin{equation}
\label{slidp:Sss}
\begin{split}
S[\;x,\; \lambda,\;q\;]&=\frac{1}{5}\;  \left(2 x_0^2-2 x_0 \lambda_0+3 \lambda_0^2\right)\\ &+ \int_{0}^{t_1}dt\;\left(\frac{1}{4}\;[\dot{x}+V'(x,\lambda)]^2+\frac{1}{4}\;[\dot{\lambda}+\lambda]^2\right)+\frac{iq}{2}\;W_d[x,\lambda]-\frac{iq}{2}\; W_d.
\end{split}
\end{equation}
Notice that the action in Eq.\ \eqref{slidp:Sss} is again quadratic. Therefore one can infer that in the asymptotic regime, $P(W_d)$ must again be of the form (Section \ref{Remarks}):
\begin{equation}
\label{slidp:ssasymp}
P(W_d)= 2\times \sqrt{\frac{\beta}{4 \pi\;d_{0}^2}}\times\sqrt{\frac{\det \textbf{A}_{\;iq=0\;}}{\det \textbf{A}^\prime_{iq=i\tilde{q}^*}}}\times e^{\;\beta\;\frac{i\tilde{q}^*}{2}\;  W_d  }\;(1+O(1/\beta)).
\end{equation}
It can be verified that the functional operator that determines the optimal trajectory, as well as the fluctuations around the optimal trajectory, is again \textbf{A}, as in the previous Sections \ref{pre} and \ref{pre2}. The difference comes in the boundary conditions; matrices $M$ and $N$ get modified accordingly.\par
\begin{figure}[htb!]
 \centering
 \includegraphics[scale=0.5]{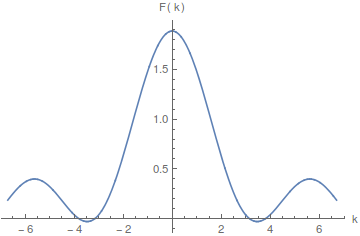}
 \caption{\label{figuresix}\textbf{$F(k)$ vs. $k$ for $T=1$}. The smallest positive and negative roots of $F(k)$ give the exact asymptotic form for the positive and negative tails of $P(W_d)$.}
\end{figure}

 As before, the leading-order behaviour of $P(W_d)$ can be found from the smallest roots of the corresponding characteristic polynomial $F(k)$ ( Eq.\ \eqref{slidp:fk} ). In Fig. \ref{figuresix} is a plot of the characteristic polynomial for $T=1$.
Notice that $F(k)$ is again symmetric about $k=0$, and this indicates that the dissipated work given by Eq.\ \eqref{slidp:ep} indeed satisfies the fluctuation relation
\begin{equation}
P(W_d)=e^{\beta W_d}\;P(-W_d).
\end{equation}
The smallest roots of $F(k)$, are found to be $k^*=\pm 3.16$ and therefore the leading-order asymptotic form is given by (for $\beta =1$):
\begin{align}
P(W_d^+)&\sim e^{-1.08199 \vert W_d \vert},& P(W_d^-) &\sim e^{-2.08199 \vert W_d \vert},
\end{align}
where the $\pm$ denotes the positive and negative tails respectively. As done earlier, one can also improve this estimate by including the pre-exponential factor. The steps involved are identical to the previous cases considered. The quadratic nature of the augmented action in Eq.\ \eqref{slidp:Sss} leads to a sub-leading power-law behaviour $\sim \vert W_d \vert ^{-\frac{1}{2}}$ , and a numerical factor which is completely determined in terms of the zero-mode of the operator \textbf{A}, and  depends only on the time duration $T$ of the protocol. Explicit calculations may be carried out in the same way as in the previous case (see Appendix \ref{ec}). For $T=1$ we find:
\begin{align}
P(W_d^+)&\sim \frac{1.13}{\sqrt{\vert W_d \vert }}\; e^{-1.08199 \vert W_d \vert},& P(W_d^-) &\sim \frac{1.13}{\sqrt{\vert W_d \vert }}\;e^{-2.08199 \vert W_d \vert}.
\end{align}
In Table \ref{tab: Asymptotic forms.} we present the exact asymptotic forms of $P(W_d)$ for different values of $T$\par
\begin{table}[h!]
  \centering
  \label{tab: Asymptotic forms.}
  \begin{tabular}{ccc}
    \toprule
    T & $P(W_d^+)$ & $P(W_d^-)$\\
    \midrule
    \vspace{2mm}
    1 & $\frac{1.13}{\sqrt{\vert W_d \vert}} \;e^{-1.08  \vert W_d\vert}$ & $\frac{1.13}{\sqrt{\vert W_d \vert}} \;e^{-2.08  \vert W_d\vert}$\\\vspace{2mm}
    2 & $\frac{3.30}{\sqrt{\vert W_d \vert}} \;e^{-0.88  \vert W_d\vert}$ &$\frac{3.30}{\sqrt{\vert W_d \vert}} \;e^{-1.88  \vert W_d\vert}$\\\vspace{2mm}
    3 & $\frac{7.44}{\sqrt{\vert W_d \vert}} \;e^{-0.78  \vert W_d\vert}$ &$\frac{7.44}{\sqrt{\vert W_d \vert}} \;e^{-1.78  \vert W_d\vert}$\\\vspace{2mm}
    4 & $\frac{10.44}{\sqrt{\vert W_d \vert}} \;e^{-0.72  \vert W_d\vert}$ &$\frac{10.44}{\sqrt{\vert W_d \vert}} \;e^{-1.72  \vert W_d\vert}$\\
    \bottomrule
       \end{tabular}
         \caption{ \textbf{Asymptotic Forms for $P(W_d,\;T)$}. The exact asymptotic forms of $P(W_d)$ for different values of $T$, obtained using Eq.\ \eqref{slidp:ssasymp}.}
\end{table}\par
\subsubsection{Comparison with results in    \cite{Varley}.}
\label{comp}
\begin{figure}[htb!]
 \centering
 \includegraphics[scale=0.35]{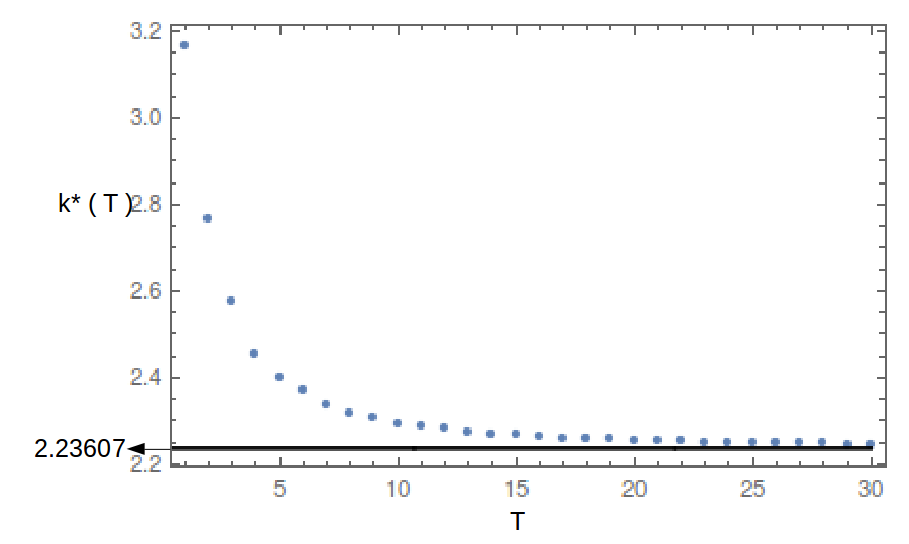}
 \caption{\label{figureseven}\textbf{$k^*(T)$ vs $T$}. In the large $T$ limits the result obtained from Eq.\ \eqref{slidp:ssasymp} for the leading-order asymptotic form of $P(W_d)$ agrees with the results in    \cite{Varley}}
\end{figure}

Verley {\it et al}    \cite{Varley}, obtain the exact form of the generating function of the probability distribution of the dissipation function for large values of the time duration $T$ of the protocol, and show that the Crooks fluctuation theorem is satisfied in this limit   \cite{Varley}. In order to compare the two methods, we have first inverted the generating function from    \cite{Varley} (details in Appendix \ref{com}). The exact form of $P(W_d,T)$ for $T\gg 1$ is obtained as :
\begin{equation}
\label{slidp:varley}
\begin{split}
P(W_d,T) &\sim  \frac{4\ 5^{3/4} \left(T^2+W_d^2\right) }{\sqrt{\pi } T \left(\frac{T}{\left(\frac{T^2}{T^2+W_d^2}\right)^{3/2}}\right)^{3/2} \left(\sqrt{5} \sqrt{\frac{T^2}{T^2+W_d^2}}+2\right){}^2}\\ &\times \exp  \left(\frac{1}{2} T \left(\frac{W_d \left(-T W_d \sqrt{\frac{5 W_d^2}{T^2}+5}+T^2+W_d^2\right)}{T \left(T^2+W_d^2\right)}-\sqrt{5} \sqrt{\frac{T^2}{T^2+W_d^2}}+2\right)\right).
\end{split}
\end{equation}
In order to compare this result with our calculations, we do an asymptotic expansion of Eq.\ \eqref{slidp:varley} for large $W_d$. To leading-order we find that,
\begin{align}
\label{slidp:varley asym}
P(W_d^+) &\sim e^{\left(\frac{1}{2}-\frac{\sqrt{5}}{2}\right) W_d}, & P(W_d^-)&\sim e^{-\left(\frac{1}{2}+\frac{\sqrt{5}}{2}\right) \vert W_d \vert }.
\end{align}  
We compare this with the leading-order form in Eq.\ \eqref{slidp:ssasymp} by checking how $k^*(T)$ behaves as $T$ becomes large. Knowing the exact form $F(k)$ for any value of $T$, this behaviour may be readily found. As we show in Fig. \ref{figureseven}, we find,
\begin{equation}
\label{slidp:compare1}
k^* \equiv 1-iq_\pm^* \xrightarrow{\text{large T}}\pm\sqrt{5},
\end{equation}
and this leads to the asymptotic forms given in Eq.\ \eqref{slidp:varley asym}. Therefore in the large-$T$ limit, the leading-order behaviour predicted by both methods agree. \par
\begin{figure}[htb!]
 \centering
 \includegraphics[scale=0.35]{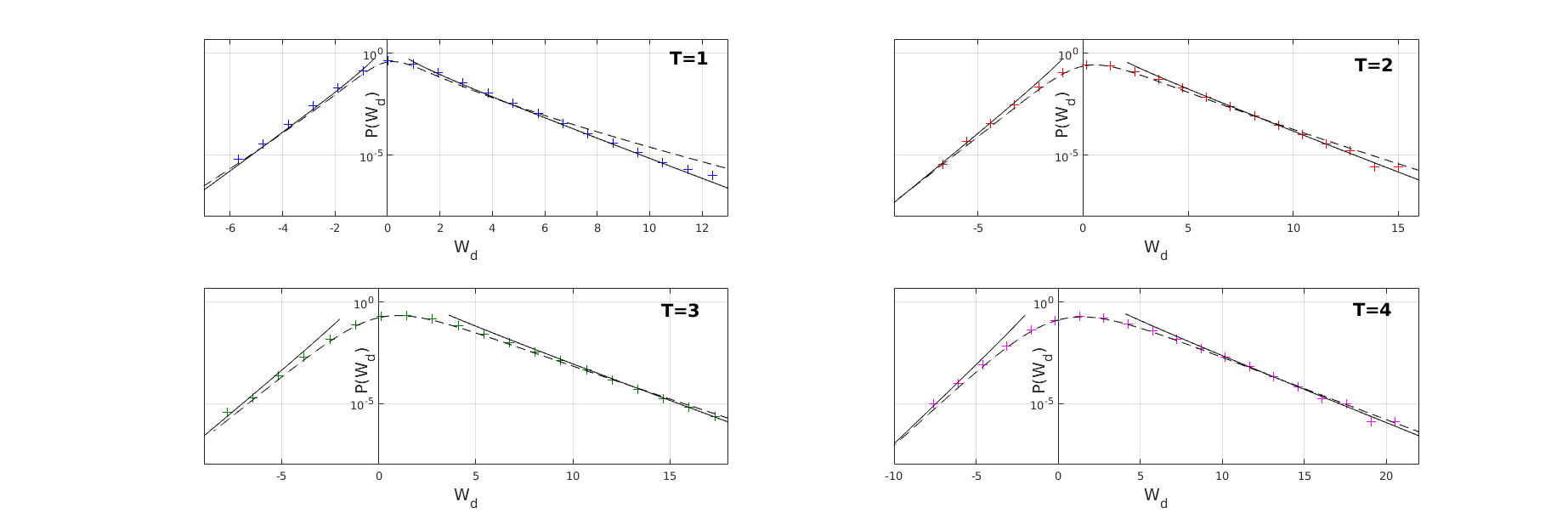}
 \caption{\label{figure8}\textbf{ Comparison of asymptotic forms obtained using Eq.\ \eqref{slidp:ssasymp}, Eq.\ \eqref{slidp:varley} and numerical simulations}. The symbols correspond to the results from the numerical simulation of the Langevin dynamics with a step size of $\Delta t=0.001$, $\beta=1$ and an average over $10^6$ realizations. The solid lines correspond to the asymptotic form obtained using Eq. \eqref{slidp:ssasymp} (Table \ref{tab: Asymptotic forms.}). Dashed lines correspond to the asymptotic form of Eq.\ \eqref{slidp:varley}.}
\end{figure}

Now we look at the sub-leading pre-exponential behaviour predicted by both methods. The asymptotic expansion of the pre-exponential factor of $P(W_d,T)$ in Eq.\ \eqref{slidp:varley} gives a sub-leading pre-exponential behaviour $\sim \vert W_d \vert ^{-5/2}$, which is a much faster decay than predicted by Eq.\ \eqref{slidp:ssasymp} which suggests a sub-leading behaviour $\sim \vert W_d \vert ^{-1/2}$ for any value of $T$. In Figure \ref{figure8}, we compare the results from both methods with our simulations of the Langevin dynamics.
The asymptotic forms obtained in Table \ref{tab: Asymptotic forms.} are good fits to the tails for all values of $T$. As $T$ gets larger, Eq.\ \eqref{slidp:varley} becomes a good fit to the numerical data, and the two methods agree to a good extent at the tails (as expected from the same leading-order behaviour). In order to see the difference coming from the disagreement in the pre-exponential factors of both methods, extensive Langevin simulations will be needed. 
\section{Conclusion.}
\label{con}
In this paper we have determined the exact asymptotic form of the work distribution, including the pre-exponential factor, in a class of stochastically driven systems, using a theory developed by Engel and Nickelsen (EN theory)    \cite{Nickelsen2011}. In cases where the exact finite time work distribution is not known, EN theory can be applied to obtain good analytical approximations for the tails of the distributions, which are generically hard to observe in experiments or numerical simulations. This can then be combined with data from experimentally / numerically viable regimes to construct the full probability distribution. The extension of EN theory to stochastically driven systems has not been done previously and the analytic solutions that we obtain here are new.\par EN theory involves writing an augmented action which carries all the information about  initial conditions as well as the functional - $W[x(\cdot)]$ whose probability distribution is to be computed. The asymptotic form of work distributions are then computed by using the saddle point approximation, which formally corresponds to the small noise limit. For a class of quadratic (augmented) actions, we have shown that the smallest roots of a certain characteristic polynomial function $F(k)$ determines both the leading-order asymptotic form as well as the pre-exponential factor. The Crooks fluctuation theorem is then shown to manifest itself as the reflection symmetry property of this function. These features can be explained using the relation (proved in Appendix \ref{ofk}),
\begin{align}
\label{mgft}
\left\langle e^{-\frac{iq}{2}\; W\left[x(\cdot)\right]}\right\rangle_T &=\sqrt{\frac{F(1)}{F(k)}},& k &\equiv 1-iq,
\end{align}
for the exact moment generating function (MGF). For a colloidal particle in a harmonic trap, where the mean position of the trap is modulated according to the Ornstein-Uhlenbeck process, we have shown that the (dissipated) work distributions have the asymptotic behaviour,
\begin{equation}
\label{univ}
P(W) \sim \frac{C_1}{\sqrt{ \vert W \vert }}\;e^{-C_2 \vert W \vert},
\end{equation}
in both the transient and the steady state. Here the constants, $C_1$ and $C_2$ are fixed by the time duration of the driving and the noise coefficients, and are explicitly determined for all cases. The asymptotic form given by Eq.\ \eqref{univ} can be shown to be universal for quadratic augmented actions. A rigorous discussion of this point and also the relation of the asymptotic form to the singularities of the MGF may be found in    \cite{linDiff,Park}. In    \cite{linDiff}, the authors have shown that for the same class of systems for which the above discussions apply, determination of various PDFs and MGFs reduce to finding solutions of certain nonlinear differential equations (NLDEs), which in many cases need to be solved numerically. The method of \textit{functional determinants} simplifies this problem to instead determining the solutions of ELEs, which are linear ordinary differential equations. In    \cite{loghar}, for the special case of a Brownian particle in a logarithmic harmonic potential, the authors obtained the asymptotic form of work distribution in terms of the solution of a Riccati differential equation. We are not aware of any other analytic methods for performing the exact finite-time computation of the asymptotic form including the pre-factor, in Langevin systems. Although we have restricted ourselves to the computation of asymptotics of work distributions in this work, Eq.\ \eqref{mgft} contains more information, such as the complete moment hierarchy. These aspects will be discussed in a future publication. \par
We believe that the methods that we discuss here has potential applications in the context of finite time thermodynamics of stochastic systems. It should be interesting to look into more applications of this theory, particularly in other stochastic potentials typical for experimental situations. For example, the theory can be developed to include stochastic driving governed by discrete time stochastic processes such as the one studied in \cite{unc}, by using an appropriate path integral representation \cite{Doi,Doii,peliti}. However, identifying universality classes for asymptotics of work distributions still remains an open problem.   
\section*{Acknowledgement}
We would like to thank Daniel Nickelsen for very helpful discussions, comments and a critical reading of an earlier version of this manuscript. We would also like to thank Viktor Holubec and Dominik Lips for pointing out an error in the references, and Gatien Verley
for helpful comments on reference    \cite{Varley}.
\section*{Author contribution}
Both the authors have contributed equally to this manuscript. 
\newpage
\appendix
\section{The Jacobian}
\label{jac}
In this Appendix, we obtain the exact form of the Jacobian of the transformations that is required in deriving Eq.\ \eqref{slidp:preexp} in Section \ref{slidp:tr}. We generalise the derivation of Engel and Nickelsen in    \cite{Nickelsen2011}, to the SSP studied in Section \ref{slidp} described by the Langevin equations,
\begin{equation}
\label{jac:system}
\begin{split}
\dot{x}(t) &= \lambda(t)-x(t)+\sqrt{\frac{2}{\beta}}\; \eta (t),\\
\dot{\lambda}(t) &= -\lambda(t)+\sqrt{\frac{2}{\beta}}\; \xi (t).
\end{split}
\end{equation}
The propagator of the corresponding Fokker-Planck equation may be written down as,
\begin{equation}
\begin{split}
p(x_T,\lambda_T,T\vert
x_0,\lambda_0,0)&=\textbf{N}\int_{x_0,\lambda_0}^{x_T,\lambda_T}\textbf{D}\left[x(\cdot),\lambda(\cdot) \right]\\ &\times\text{exp}\left(\;-\frac{\beta}{4}\int_0^T dt\;\left(\left(\dot{x}+x-\lambda \right)^2+\left(\dot{\lambda}+\lambda \right)^2\right)\right).
\end{split}
\end{equation}
From the normalization condition 
\begin{equation}
\frac{1}{Z_0}\sqrt{\frac{\beta}{2\pi}}\int dx_T\;\int dx_0\;\int d\lambda_T\;\int d\lambda_0\;  e^{-\beta \tilde{V}_0} \; p(x_T,\lambda_T,T\vert
x_0,\lambda_0,0)=1,
\end{equation}
where,
\begin{equation}
\tilde{V}_0=\frac{(x_0-\lambda_0)^2}{2}+\frac{\lambda_0^2}{2},
\end{equation}
we have,
\begin{equation}
\label{jac:nor}
\begin{split}
1 &=\frac{\textbf{N}}{Z_0}\sqrt{\frac{\beta}{2\pi}}\int dx_T\;\int dx_0\;\int d\lambda_T\;\int d\lambda_0\;  e^{- \beta \tilde{V}_0} \int_{x_0,\lambda_0}^{x_T,\lambda_T}\textbf{D}\left[x(\cdot),\lambda(\cdot) \right]\\ &\times  \text{exp}\left(\;-\frac{\beta}{4}\int_0^T dt\;\left(\left(\dot{x}+x-\lambda \right)^2+\left(\dot{\lambda}+\lambda \right)^2\right)\right).
\end{split}
\end{equation}
After several partial integrations in the RHS of Eq.\ \eqref{jac:nor} we obtain,
\begin{equation}
\label{jac:g}
\begin{split}
1 &= \frac{ \textbf{N}}{Z_0}\sqrt{\frac{\beta}{2\pi}}\int dx_T\;\int dx_0\;\int d\lambda_T\;\int d\lambda_0\;  e^{- \beta \tilde{V}_0} \; \int_{x_0,\lambda_0}^{x_T,\lambda_T}\textbf{D}\left[\begin{array}{c c} x(\cdot)& \lambda(\cdot) \end{array}\right]\\ &\times \text{exp}\left(\;-\frac{\beta}{4}\int_0^T dt\;\left[\begin{array}{cc}x&\lambda
\end{array} \right] \;A\;\left[\begin{array}{c}
x\\
   \lambda\\
\end{array} \right]
    \;+\left(\;x \dot{x}+x^2 -\lambda x +\lambda \dot{\lambda}+\lambda^2 \; \right) \bigg\vert_0^T \right).
\end{split}
\end{equation}
The above Gaussian integral is straightforward to compute and gives,
\begin{equation}
1 = \frac{\textbf{J}\;\textbf{N}}{Z_0}\sqrt{\frac{\beta}{2\pi}}\;\frac{1}{\sqrt{\det \textbf{A}}},
\end{equation}
where \textbf{J} is the Jacobian to be determined. Note that $\textbf{A} = \textbf{A}_{iq=0}$ as given by Eqns.\ \eqref{slidp:ele} and \eqref{slidp:elebc}. Therefore we have,
\begin{equation}
\textbf{J}= \frac{Z_0}{\textbf{N}} \sqrt{\frac{2\pi}{\beta}}\times \sqrt{\det \textbf{A}_{\;iq=0\;}}.
\end{equation}
This is the result used in deriving Eq.\ \eqref{slidp:preexp}. In a similar manner one can show that all the other Jacobians appearing in the main text, have the above form.
\section{Functional determinants}
\label{fdet}
The necessity of computing functional determinants of certain differential operators arises in many different situations. As we have seen in the main text, computing the leading-order contribution to path integrals is one of them. In many cases it is not an absolute functional determinant that is required, but a ratio where one or both of the operators can in principle have a zero-mode. Profound mathematical techniques have been developed to compute functional determinants (or the ratio of determinants) even in situations when the operators have zero-modes. Here we apply the contour integration method suggested in    \cite{Kirsten} to the SSP problem discussed in Section \ref{slidp}. As we have already seen, the operators that appear in the ratio of determinants in Eq.\ \eqref{slidp:preexp} are $2\times 2$ matrix differential operators. In    \cite{Kirsten} Kirsten {\it et al}, have discussed the possible generalization of their techniques to such matrix differential operators as well. Recently in    \cite{functional}, Falco {\it et al}  have also looked at a similar generalization. In contrast to these previous studies, the operator \textbf{A} that we study here has differential operator entries in the off diagonal terms as well. However, the methods discussed in    \cite{Kirsten} can also be generalized to this situation. In the case of the SSP, the matrix differential operator that we need to find the determinant of, is defined by the following problem:
\begin{align}
\label{zeta:eql}
A\left[\begin{array}{c}
x(t)\\
\lambda(t)\\
\end{array} \right]&=l \left[\begin{array}{c}
x(t)\\
\lambda(t)\\
\end{array} \right], & \text{ where } \textbf{A} &= \left[\begin{array}{cc}-\frac{d^2}{d 
t^2}+1 & k\;\frac{d}{d t}-1\\-k\;\frac{d}{d t}-1 & -\frac{d^2}{d 
t^2}+2 \end{array} \right];\; k\equiv 1-iq,
\end{align}
together with Robin-type boundary conditions:
\begin{align}
\label{zeta:bc}
M\left[
\begin{array}{c}
x(0)\\
\lambda(0)\\
\dot{x}(0)\\
\dot{\lambda}(0)\\
\end{array}
\right]&=0, & N\left[
\begin{array}{c}
x(T)\\
\lambda(T)\\
\dot{x}(T)\\
\dot{\lambda}(T)\\
\end{array}
\right]&=0.
\end{align}
The form of the matrices $M$ and $N$ can be deduced from Eq.\ \eqref{slidp:elebc}. Using the results from    \cite{Kirsten}, one can then write down the determinant ratio as,
\begin{align}
\frac{\det \textbf{A}_{\;iq=0\;}}{\det \textbf{A}^\prime_{\;iq=\i\tilde{q}^*\;}} =\frac{\det\left[M+NH_{k=1}(T)\right]}{\textbf{B}\langle\textbf{u}_N(t)\vert \textbf{u}_N(t) \rangle}=\frac{F(1)}{\textbf{B}\langle\textbf{u}_N(t)\vert \textbf{u}_N(t) \rangle}.
\end{align}
Here $H_k$ is the matrix of fundamental solutions of the homogeneous equation,
\begin{equation}
A\left[\begin{array}{c}
x(t)\\
\lambda(t)\\
\end{array} \right]=0,
\end{equation}
defined as,
\begin{align}
\label{eqhkt}
H_k(t) &=\left[
\begin{array}{cccc}
 x_1(t) & x_2(t) & x_3(t) & x_4(t) \\
 \lambda_1(t) & \lambda_2(t) & \lambda_3(t) & \lambda_4(t)\\
 \dot{x}_1(t) & \dot{x}_2(t) & \dot{x}_3(t) & \dot{x}_4(t)\\
\dot{\lambda}_1(t) & \dot{\lambda}_2(t) & \dot{\lambda}_3(t) & \dot{\lambda}_4(t) \\
\end{array}
\right],& H_k(0)=\textbf{I}_4.
\end{align}
The function \textbf{B} appear due to the presence of the zero-mode and need to be determined using the self adjointness property of the differential operator \textbf{A} in each case. The normalized zero-mode, $\textbf{u}_N(t)$ is defined as,
\begin{equation}
\begin{split}
\textbf{u}_N(t)=\left[
\begin{array}{c}
x_N(t)\\
\lambda_N(t)\\
\end{array}
\right] &= x(0)\left[
\begin{array}{c}
x_1(t)\\
\lambda_1(t)\\
\end{array}
\right]+\lambda(0)\left[
\begin{array}{c}
x_2(t)\\
\lambda_2(t)\\
\end{array}
\right]+\dot{x}(0)\left[
\begin{array}{c}
x_3(t)\\
\lambda_3(t)\\
\end{array}
\right]\\&+ \dot{\lambda}(0)\left[
\begin{array}{c}
x_4(t)\\
\lambda_4(t)\\
\end{array}
\right],
\end{split}
\end{equation}
where the constants are determined by,
\begin{equation}
\label{norm:ini}
\left[
\begin{array}{c}
x(0)\\
\lambda(0)\\
\dot{x}(0)\\
\dot{\lambda}(0)\\
\end{array}
\right]=\text{Adjoint}\left[M+NH_{k}(T)\right]\left[
\begin{array}{c}
0\\
0\\
0\\
1\\
\end{array}
\right].
\end{equation}
The inner product is the usual one, given by
\begin{equation}
\langle\textbf{u}_N(t)\vert \textbf{u}_N(t) \rangle=\vert\vert\left[
\begin{array}{c}
x_N(t)\\
\lambda_N(t)\\
\end{array}
\right]\vert\vert^2=\int_0^Tdt\;\left( x_N^2(t)+\lambda_N^2(t)\right).
\end{equation}
In case of the SSP discussed in Section \ref{slidp:tr}, we find, 
\begin{equation}
\textbf{B}=\frac{1}{\lambda_N(T)}
\end{equation}
In the next Section, we show how this theory can be used to compute the functional determinants that appear in the main text.
\section{Explicit Computations}
\label{ec}
Here we provide the explicit calculations using EN theory for two of the cases considered in the main text, the breathing parabola in Section \ref{brethp} and the SSP in Section \ref{slidp:tr}. For the breathing parabola problem, in    \cite{Nickelsen2011}, for a specific choice of (reverse) protocol, EN theory was used to compute the exact asymptotic form of $P(W)$. Here we present the calculations for a particular choice of forward protocol, and give only the final solution for the corresponding reverse protocol. The solutions we obtain for the SSP problem in Section \ref{SSP} are new, and are generalizations of the calculations in Section \ref{BP}.
\subsection{Breathing Parabola: $P_{F/R}(W)$}
\label{BP}
For the breathing parabola, we have considered the specific forward protocol,
\begin{align}
\lambda(t)&=\frac{1}{2-t},& t&=0\text{ to }1.
\end{align}
In terms of the shifted variable $k=1-iq$, the corresponding ELE reads (for simplicity we will use $x(t)$ instead of $\tilde{x}(t)$ everywhere.),
\begin{equation}
\ddot{x}(t)+\left(\frac{k}{(t-2)^2}-\frac{1}{(t-2)^2}\right) x(t)=0.
\end{equation}
Two independent solutions of this 2nd order differential equations are given by,
\begin{align}
\label{bretp:funda}
x_1(t)&=(t-2)^{\frac{1}{2} \left(1-\sqrt{5-4 k}\right)} ,& x_2(t)&=(t-2)^{\frac{1}{2} \left(1+\sqrt{5-4 k}\right) }.
\end{align}
Together with the boundary conditions,
\begin{align}
M\;\left[\begin{array}{c}
x_0\\\dot{x}_0
\end{array}\right] &= 0,& N \left[\begin{array}{c}
x_T\\\dot{x}_T
\end{array}\right]&=0.
\end{align}
where M and N are matrices,
\begin{align}
M&=\left[\begin{array}{c c}\lambda_0 & -1 \\0&0 
\end{array}\right],& N&=\left[\begin{array}{cc}0&0\\\lambda_T &1 
\end{array}\right].
\end{align}
The above system constitutes a second order Sturm-Liouville eigenvalue problem in $k$. A non trivial solution exists only for some specific values of $k$, which are given by the roots of the corresponding characteristic  polynomial,  
\begin{equation}
F(k)=\det\left[M+N H_k(T)\right].
\end{equation}
As we have seen in the main text, the asymptotic behaviour of $P(W)$ is determined by the smallest value of $k(\equiv k^*)$ for which $F(k)=0$. The value of $k^*$ may be obtained numerically. For the case $T=1$ we find $k^*=3.67$ (see Figure \ref{figureone} in the main text). The leading-order asymptotic form of $P(W)$ is therefore,
\begin{equation}
P_F(W)\sim e^{-1.33 \;\vert W\vert}.
\end{equation}
In order to improve this estimate one has to compute the pre-exponential factor as well. The computation of the factor $d_0^2$ is rather straightforward. From \eqref{bretp:dn} we see that,
\begin{equation}
\label{brethp:d0}
d_0 = \frac{2 W}{\vert \vert x(t) \vert \vert},
\end{equation}
where $x(t)$ is the zero-mode. For $k=-3.67$, we find that the zero-mode is 
\begin{equation}
x(t)=C_1\; (t-2)^{0.5 -1.55\; i} \left((t-2)^{3.11\; i}-\frac{\left(t+1+(-0.5-1.55\; i)\right) (t+1)^{3.11\; i}}{ t+1+(-0.5+1.55\;i)}\right).
\end{equation}
Here $C_1$ is the undetermined constant in the solution, which is to be fixed using the constraint equation \eqref{bretp:constraint}. Using the above form of $x(t)$ and the constraint equation \eqref{bretp:constraint}, we find,
\begin{equation}
d_0 = \frac{2 W}{\vert \vert x(t) \vert \vert}=0.97\; \sqrt{\vert W \vert}.
\end{equation}
The other factor which appears in the calculation of the pre-exponential factor is the determinant ratio of the two functional differential operators,
\begin{equation}
\label{bretp:detra}
\sqrt{\frac{\det \textbf{A}_{\;iq=0\;}}{\det\textbf{A}^\prime_{\;iq=i\tilde{q}^*\;}}}=\sqrt{\frac{\det\left[M+N\;H(1)\; \right]_{iq=0}}{\textbf{B} \langle x_{N}(t) \vert x_{N}(t) \rangle}}.
\end{equation}
Let us first compute the factor appearing in the numerator of the RHS. Using Eq.\ \eqref{bretp:funda}, the matrix of normalized fundamental solutions ($ H(0)=\textbf{I}_2$) when $iq=0$ may be found as,
\begin{equation}
H(t)=\left[
\begin{array}{cc}
 1 & t \\
 0 & 1 \\
\end{array}
\right].
\end{equation}
Therefore,
\begin{equation}
\det\left[M+N\;H(1)\; \right]_{iq=0}=-2.
\end{equation}
Notice that, this is also the limiting value of $F(k)$ as $k\rightarrow 1$ in Figure \ref{figureone}. Let us now look at the term in the denominator of the RHS of Eq.\ \eqref{bretp:detra}. The appropriately normalized solutions $x^{N}(t)$ can be identified using \eqref{bretp:funda} and Eq.\ \eqref{norm:ini} adapted to this problem. We obtain, 
\begin{equation}
x_N(t)=(-0.0011 + 0.0035\; i) (-2 + t)^{( 0.5 - 1.55\;i)} + (19.92 + 61.97\;i) (-2 + t)^{(0.5 + 1.55\;i)}.
\end{equation}
Using the methods discussed in    \cite{Kirsten} we find that for this problem,  $\textbf{B}=-\frac{1}{x_N(T)}$. Together with this, one finds
\begin{equation}
\textbf{B}\; \langle x_{N}(t) \vert x_{N}(t) \rangle=-1.26.
\end{equation}
Putting the factors together in Eq.\ \eqref{bretp:fullpw}, we finally get,
\begin{equation}
P_F(W)\approx \frac{0.73}{\sqrt{\vert W \vert}}\;e^{-1.33 \vert W\vert}.
\end{equation}
Similarly for the reverse protocol,
\begin{align}
\lambda(t)&=\frac{1}{1+t}, & t&=0\text{ to }1,
\end{align}
it can be shown that,
\begin{equation}
P_R(W)\approx \sqrt{2}\times \frac{0.73}{\sqrt{\vert W \vert}}\;e^{-2.33 \vert W\vert}.
\end{equation}
We compare these results with numerical simulations, and as we show in the main text, Figure \ref{figuretwo}, the prediction for the tail region is in excellent agreement with the theoretical predictions.
\subsection{The Stochastic Sliding Parabola}
\label{SSP}
In this Appendix, we do explicit computations to obtain the 
asymptotic form of the probability distribution Eq. \eqref{eq:FF_SSP}, found in Section \ref{slidp:tr} for equilibrium initial conditions and $T=1$ (explicit calculations for the other cases discussed in section \ref{free} and section \ref{slip:ss} can be carried out in a similar manner). For simplicity we will use the notation $x(t)$ and $\lambda(t)$ instead of $\tilde{x}(t)$ and $\tilde{\lambda}(t)$. \par First we note that when $\vert k\vert \neq1$, following some algebra, the system of Euler-Lagrange equations for $\left(x,\lambda \right)$ given by Eq.\ \eqref{slidp:ele} in the main text can be reduced to a fourth order ordinary differential equation for one of the variables (for example, $\lambda$) as,
\begin{equation}
\label{slip:ele}
\ddddot{\lambda}+(k^2-3)\ddot{\lambda}+\lambda(t)=0.
\end{equation}
In terms of the solution $\lambda(t)$, $x(t)$ is then given by,
\begin{equation}
x(t)=\frac{\left(k^2-2\right) \lambda(t)+k \left(2-k^2\right) \dot{\lambda}(t)-k \dddot{\lambda}(t)+\ddot{\lambda}(t)}{k^2-1}.
\end{equation}
Eq.\ \eqref{slip:ele} has four independent solutions given by,
\begin{equation}
\lambda(t)=e^{\pm\frac{\sqrt{3-k^2\pm\sqrt{k^4-6 k^2+5}}}{\sqrt{2}}\;t}.
\end{equation}
A general solution for a specific optimal trajectory ($k=k^*$) can always be written as a linear combination of these four independent solutions, where the coefficients are fixed by the boundary conditions and the constraint equation. As we discussed in Section \ref{slip:lead}, in order to compute the leading-order behaviour of $P(W)$, we only require the $k^*$ values and not the explicit solution. In order to find $k^*$, we look at the smallest roots of the function: 
\begin{equation}
F(k)=\det\left[\;M+N\;H_k(T)\right].
\end{equation}
$M$ and $N$ corresponding to the boundary conditions in \eqref{slidp:elebc} can be written down as, 
\begin{align}
M&=\left[\begin{array}{cccc}
 -k-1 & k+2 & 0 & -1 \\
 1 & -1 & -1 & 0 \\
 0 & 0 & 0 & 0 \\
 0 & 0 & 0 & 0 \\
\end{array}
\right], & N &=\left[
\begin{array}{cccc}
 0 & 0 & 0 & 0 \\
 0 & 0 & 0 & 0 \\
 1 & -1 & 1 & 0 \\
 k-1 & 2-k & 0 & 1 \\
\end{array}
\right].
\end{align}
$H_k(t)$ has the form as given in Eq.\ \eqref{eqhkt}. From Figure \ref{figurethree} in the main text, we see that the relevant $k^*$ values are given by $k^*=\pm 2.3$. Solving the ELEs \eqref{slidp:ele} along with the boundary conditions \eqref{slidp:elebc}, for $k =2.3$ yields,
\begin{align}
\label{slip:op1}
\lambda(t)&=C_1 \;(0.0043 \sin (0.76 t)-1.30 \sin (1.30 t)-0.0056 \cos (0.76 t)+ \cos (1.30 t)),\\
x(t)&=C_1 \;(-0.0035 \sin (0.76 t)+0.62 \sin (1.30 t)- 0.0083 \cos (0.76 t)+1.82 \cos (1.30 t)).
\end{align}
If we evaluate the work done along this optimal trajectory, we get,
\begin{equation}
W[x]=\int_0^{1}\;dt \;\left(\lambda(t)-x(t)\right)\dot{\lambda} =3.38495 \;C_1^2.
\end{equation}
The work done is positive; this indicates that this pair of trajectories correspond to the positive tail of P(W). Similarly, solving the ELEs \eqref{slidp:ele},\eqref{slidp:elebc} for $k=-2.3$ gives,
\begin{align}
\label{slip:opt2}
\lambda(t)&=C_1\; (0.0043 \sin (0.76 t)-1.30 \sin (1.30 t)-0.0056 \cos (0.76 t)+1. \cos (1.30 t)),\\
x(t)&=C_1\; (0.0089 \sin (0.76 t)-1.59 \sin (1.30 t)+ 0.0012\cos (0.76 t)-1.08 \cos (1.30 t)).
\end{align}
The work done along this trajectory become, 
\begin{equation}
W[x]=\int_0^{1}\;dt \;\left(\lambda(t)-x(t)\right)\dot{\lambda} =-3.38495 \;C_1^2.
\end{equation}
This value is negative, and therefore $k^*=-2.3$ corresponds to the negative tail of $P(W)$. With this we find that to leading-order, the positive and negative tails of P(W) have the functional form,
\begin{align}
P(W^+)&\sim e^{-0.65\;\vert W\vert},& P(W^-)&\sim e^{-1.65\vert W \vert},
\end{align}
respectively. \par In order to improve this estimate, we next include the pre-exponential factor. The first factor which goes into the pre-exponential is $d_0^2$ defined in Eq.\ \eqref{slidp:d0}. This can be computed relatively easily as in the case of the breathing parabola. For both $k=\pm\; 2.3$ we find using Eq.\ \eqref{slip:op1} and \eqref{slip:opt2},
\begin{equation}
d_0^{\;\pm}=\frac{2\; \vert W\vert}{\vert\vert \left[\begin{array}{c}x\\ \lambda 
\end{array} \right]^\pm\vert\vert }=2.032 \;\sqrt{\vert W\vert}.
\end{equation}
(The superscript  $\pm$ is used to denote the solutions for positive or negative tails). The next factor to be computed is the square root of the ratio of two functional determinants, for which we will use the formula,
\begin{equation}
\label{rhs}
\sqrt{\frac{\det \textbf{A}_{\;iq=0\;}}{\det \textbf{A}^\prime_{iq=iq^*_\pm}}}=\sqrt{\frac{F(1)}{\textbf{B} \langle\textbf{u}_N^{\;\pm}(t) \vert \textbf{u}_N^{\;\pm}(t) \rangle}},
\end{equation}
where,
\begin{equation}
\label{slip:detra}
\textbf{u}_N^\pm(t)=\left[\begin{array}{c}x_N(t)\\\lambda_N(t)
\end{array}\right]^\pm.
\end{equation}
is the appropriately normalized solution to the ELEs, which have to be found using Eq.\ \eqref{norm:ini}. For this particular case, We find that,
\begin{equation}
\left[\begin{array}{c}x_N(t)\\\lambda_N(t)
\end{array}\right]^+= \left[\begin{array}{c}-0.0011 \sin (0.76 t)+0.20 \sin (1.30 t)-0.0027 \cos (0.76 t)+0.59 \cos (1.30 t)\\0.0014 \sin (0.76 t)-0.42 \sin (1.30 t)-0.0018 \cos (0.76 t)+0.32 \cos (1.30 t)
\end{array}\right].
\end{equation}
Similarly
\begin{equation}
\left[\begin{array}{c}x_N(t)\\\lambda_N(t)
\end{array}\right]^-= \left[\begin{array}{c}0.023 \sin (0.76 t)-4.11 \sin (1.30 t)+0.0032 \cos (0.76 t)-2.79 \cos (1.30 t)\\0.011 \sin (0.76 t)-3.37 \sin (1.30 t)-0.014 \cos (0.76 t)+2.58 \cos (1.30 t)
\end{array}\right].
\end{equation}
Using the self-adjointness property of \textbf{A}, one can again compute, 
\begin{equation}
\textbf{B} =\dfrac{1}{\lambda_N(T)}.
\end{equation}
Also using Figure \ref{figurethree} to compute $F(1)$, we find,
\begin{equation}
\sqrt{\frac{F(1)}{\textbf{B} \langle\textbf{u}_N^{\;\pm}(t) \vert \textbf{u}_N^{\;\pm}(t) \rangle}}=1.87.
\end{equation}
Therefore the full pre-exponential factor is,
\begin{equation}
2\times \sqrt{\frac{\beta}{4 \pi\;d_{0,\pm}^2}}\times\sqrt{\frac{\det \textbf{A}_{\;iq=0\;}}{\det \textbf{A}^\prime_{iq=iq^*_\pm}}}=\frac{0.52}{\sqrt{\vert W\vert}},
\end{equation}
and the asymptotic form of $P(W)$, including the pre-exponential factor becomes,
\begin{align}
P(W^+)&\sim \frac{0.52}{\sqrt{\vert W\vert}}\; e^{-0.65\;W}, &P(W^-)&\sim\frac{0.52}{\sqrt{\vert W\vert}}\; e^{-1.65\vert W \vert}, & (T&=1.)
\end{align}
In a similar manner, the exact asymptotic forms discussed in Section \ref{free} and Section \ref{slip:ss} can be computed for any value of $T$. In particular, for the case that we discussed in Section \ref{free}, the exact asymptotic form can be obtained as a function of $T$.
\section{Origin of the symmetry of $F(k)$}
\label{ofk}
In this section we will show the relation between $F(k)$ and the exact moment generating function of dissipated work,  which explains the reflection symmetry of $F(k)$. First, using the path integral representation, an exact relation for the moment generating function can be written down as,
\begin{align}
\label{slidp:pwdipo}
\langle e^{-\frac{iq}{2}\;W_d\left[x(\cdot),\;\lambda(\cdot)\right]}\rangle_T &=\frac{\textbf{N}}{Z_0} \int dx_0 \;\int d\lambda_0\;\int dx_T\;\int d\lambda_T\;\int_{x(0),\lambda(0)=x_0,\lambda_0}^{x(T),\lambda(T)=x_T,\lambda_T}\; \mathcal{D}[x,\; \lambda]\;e^{-\beta \; S[\;x,\; \lambda,\;q\;]},\;
\end{align}
In all the cases we have considered, the augmented action $S[\;x,\; \lambda,\;q\;]$ is quadratic, therefore by doing several partial integrations, it can be shown that it reduces to
\begin{align}
S[\;x,\; \lambda,\;q\;]&=\frac{1}{4}\;\left[\begin{array}{cc}x&\lambda
\end{array}\right]\;\textbf{A}_k\;\left[\begin{array}{c}x\\\lambda
\end{array}\right]+\text{Boundary terms in } (x, \lambda, k),& k&=1-iq.
\end{align}
where the kernel $\textbf{A}_k$ is defined by the same operator that determines the optimal trajectory (Eq. \eqref{bretp:ele}, \eqref{slidp:ele}) with the same boundary terms. Therefore the integral in Eq.\ \eqref{slidp:pwdipo} is a standard Gaussian integral which can be computed as, 
\begin{align}
\label{slidp:pwdipo2}
G(\frac{iq}{2})\equiv\langle e^{-\frac{iq}{2}\;W_d}\rangle_T &=\sqrt{\frac{\det A_{\;k=1\;}}{\det A_k}}.
\end{align}
This determinant ratio can then be computed using the techniques developed in    \cite{Kirsten}. In terms of the function $F(k)$, we find,
\begin{align}
\label{fipo}
\frac{\det A_{\;k=1\;}}{\det A_k}=\frac{F(1)}{F(k)}\Rightarrow G(\frac{iq}{2}) =\sqrt{\frac{F(1)}{F(k)}}.
\end{align}
Due to Crooks fluctuation theorem    \cite{Crooks}, the moment generating function of dissipated work ($G$) must satisfy the relation,
\begin{equation}
G(\frac{iq}{2})=G(1-\frac{iq}{2}).
\end{equation}
By writing $iq$ as $1-k$ and using Eq.\ \eqref{fipo}, it can be immediately verified that for the above relation to hold, the function $F$ must satisfy, 
\begin{equation}
F(k)=F(-k).
\end{equation}
Hence the symmetry property of $F(k)$ is a consequence of the fluctuation theorem.
\section{Comparison with the results in    \cite{Varley}}
\label{com}
In this Appendix, we will invert the generating function of $P(W_d)$ obtained in    \cite{Varley} using the methods discussed in    \cite{Sabhapandit} and    \cite{sabha1}. The solution that we obtain here will be used for comparison with the asymptotic form of $P(W_d)$ calculated using the EN theory in Section \ref{slip:ss}. \par Verley {\em et al}, in    \cite{Varley}, have shown that for very large $T$, the probability generating function of the dissipated work has the form, 
\begin{align}
\label{Z1}
Z(\mu, T)&= \langle e^{\mu W_T} \rangle \xrightarrow{\text{large T}} g(\mu )\; e^{T \phi(\mu)},& \text{ where } \langle e^{\mu W_T} \rangle &= \int_{-\infty}^\infty dW_T\; e^{\mu\;W_T} \;P(W_T).
\end{align}
The notation $W_T$ stands for the dissipated work $W_d$ over a time duration $T$ of the driving. For the SSP considered in Section\ref{slip:ss}, the functions $\phi$ and $g$ are given by    \cite{Varley}, 
\begin{align}
\label{Z2}
\phi(\mu) &= 1-\nu(\mu), \text{ where } \nu(\mu)= \sqrt{1-\mu(1+\mu)}, & g(\mu) &= \frac{4 \nu(\mu)}{\left(1+ \nu(\mu)\right)^2}.
\end{align}
The probability density function can be obtained from the moment generating function $Z(\mu,t)$ by taking the inverse Fourier (two-sided Laplace) transform:
\begin{equation}
P(W_T)=\frac{1}{2 \pi i} \int_{-i \infty}^{+i \infty} Z(\mu, T)\;e^{-\mu W_T}\;d\mu,
\end{equation}
where the integration is done along the imaginary axis in the complex-$\mu$ plane    \cite{sabha1}. Using the large T form of $Z(\mu, T)$ given by Eq.\ \eqref{Z1} and  \eqref{Z2} we write,
\begin{equation}
\label{Z3}
P(W_{T}=w T)\sim \frac{1}{2 \pi i} \int_{-i\infty}^{+i \infty} g(\mu) \; e^{T f_w(\mu)}\;d\mu,
\end{equation}
where
\begin{equation}
\label{fmu}
f_w(\mu)=1-\nu(\mu)-\mu w.
\end{equation} 
The large-$T$ form of $P(W_T)$ can be obtained from Eq.\ \eqref{Z3} by using the method of steepest descent (for completeness, we reproduce the method and discussion from    \cite{sabha1} here ).  The saddle point $\mu^*$ is obtained from the solution of the condition $f_w^\prime(\mu^*)=0$ as
\begin{equation}
\label{mus}
\mu^*(w)=\frac{1}{2} \left(\frac{\sqrt{5} w}{\sqrt{w^2+1}}-1\right).
\end{equation}
From the above expression, one finds that $\mu^*(w \rightarrow ^+_- \infty)\rightarrow \mu_{\pm}$, where
\begin{equation}
\label{lim}
\mu_{\pm}=\frac{1}{2} \left(\pm\sqrt{5}-1\right).
\end{equation}
Therefore $\mu^* \in \left(\mu_-,\mu_+ \right)$. It is useful to notice that in terms of $\mu_\pm$, 
\begin{equation}
\nu(\mu)=\sqrt{\left(\mu-\mu_-\right)\left( \mu_+ - \mu \right)}.
\end{equation}
On the real axis, outside the interval $\left[\mu_-, \mu_+ \right]$, $\nu(\mu)$ is therefore imaginary. However, in order for the the integral in the definition of $Z$ (Eq.\ \eqref{Z1}) to converge, $Z(\mu,T)$ must be real for real values of $\mu$. For this reason, it is only within the range $\mu_-<\mu<\mu_+$ ( for which $\nu(\mu)$ is real and analytic), that the analytic continuation of $Z(\mu,T)$ to real $\mu$ is allowed. We hence expect the saddle point to also lie between these values. As we have already seen in Eq.\ \eqref{lim}, this is indeed the case. Since for $\mu_-<\mu<\mu_+$, $g(\mu)$ is analytic (the denominator is positive for all $\mu$ in the specified range), it can be neglected in the saddle point calculation as a sub-leading contribution.\par The saddle point calculation relates $\phi(\mu)$ to the large deviation function $h_s(w)$ by the Legendre transform,
\begin{equation}
\label{hs}
h_s(w):=f_w(\mu^*)=\frac{1}{2} \left(-\frac{\sqrt{5} w^2}{\sqrt{w^2+1}}-\sqrt{5} \sqrt{\frac{1}{w^2+1}}+w+2\right).
\end{equation}
We also see that,
\begin{equation}
f^{''}_w(\mu^*)=\frac{2}{\sqrt{5} \left(\frac{1}{w^2+1}\right)^{3/2}}>0.
\end{equation}
This means that $f_w(\mu)$ has a minimum at $\mu^*$ along real $\mu$. Now since $g(\mu)$ is analytic, the usual saddle point approximation method   \cite{Sabhapandit} gives,
\begin{equation}
\label{pf}
P(W_T=w \;T)\sim \frac{g(\mu^*)e^{T\; h_s(w)}}{\sqrt{2\pi T f^{\prime\prime}_w(\mu^*)} }.
\end{equation}
Using, Eq.\ \eqref{hs}, \eqref{mus}, \eqref{fmu} and also the relation $w=W_d/T$ in Eq.\ \eqref{pf}, we finally get
\begin{equation}
\label{pwd}
\begin{split}
P(W_d,T) &\sim  \frac{4\ 5^{3/4} \left(T^2+W_d^2\right) }{\sqrt{\pi } T \left(\frac{T}{\left(\frac{T^2}{T^2+W_d^2}\right)^{3/2}}\right)^{3/2} \left(\sqrt{5} \sqrt{\frac{T^2}{T^2+W_d^2}}+2\right){}^2}\\ &\times \exp  \left(\frac{1}{2} T \left(\frac{W_d \left(-T W_d \sqrt{\frac{5 W_d^2}{T^2}+5}+T^2+W_d^2\right)}{T \left(T^2+W_d^2\right)}-\sqrt{5} \sqrt{\frac{T^2}{T^2+W_d^2}}+2\right)\right).
\end{split}
\end{equation}
We use this form of $P(W_d)$ in Section \ref{comp} to compare with numerical results as well as the analytic forms obtained using the EN theory. 
\newpage
\bibliographystyle{unsrt}
\bibliography{Mybib}
\end{document}